\input harvmac

\def\h {{1\over 2}}
\def\al {\alpha}

\def\ov {\overline}
\def\o {\over}
\def\Li {{\cal L}i}

\def\cmp {{\it Comm. Math. Phys.\ }}
\def\np {{\it Nucl. Phys.\ } {\bf B\ }}
\def\pl {{\it Phys. Lett.\ } {\bf B\ }}
\def\pr {{\it Phys. Rept.\ } }

\def\br{\hfill\break}
\def\IZ{ {\bf Z}}
\def\IF{{\bf F}}
\def\subsubsec #1{\ \br \noindent {\it #1} \br}
\def\Ac {{\cal A}}

\def\Ic {{\cal I}}
\def\Kc {{\cal K}}
\def\Lc {{\cal L}}
\def\Nc {{\cal N}}
\def\Oc {{\cal O}}
\def\Rc {{\cal R}}

\def\th {\theta}

\def\tr {{\rm Tr}}
\def\det {{\rm det}}

\def\lf {\left}
\def\ri {\right}
\def\ra {\rightarrow}

\def\al {\alpha}
\def\re {{\rm Re}}
\def\im {{\rm Im}}
\def\p {\partial}
\def\la{\langle}
\def\ra{\rangle}

\def\eps{\epsilon}

\lref\klt{V. Kaplunoksky, J. Louis and S. Theisen, \pl {\bf 357}
(1995) 71}
\lref\kv{S. Kachru and C. Vafa, \np {\bf 450} (1995) 69}

\lref\higgs {M. Dine, P. Huet and N. Seiberg, \np {\bf 322} (1989) 301;
L. Ibanez, W. Lerche, D. L{\"u}st and S. Theisen, \np {\bf 352} (1991) 435}

\lref\msV {P. Mayr and S. Stieberger, \pl {\bf 355} (1995) 107}
\lref\jv {{\it see e.g.:} J.--P. Derendinger, S. Ferrara, C. Kounnas
and F. Zwirner, \np {\bf 372} (1992) 145;
V. Kaplunovsky and J. Louis, \np {\bf 444} (1995) 191}

\lref\bachas {C. Bachas, C. Fabre, E. Kiritsis, N.A. Obers and
P. Vanhove, hep--th/9707126}
\lref\st {S. Stieberger, NEIP--001/97, work in progress}
\lref\borch {R.E. Borcherds, {\it Invent. Math.} {\bf 120} (1995) 161} 
\lref\WL {W. Lerche, \np {\bf 308} (1988) 102}
\lref\lsw {W. Lerche, B.E.W. Nilsson and A.N. Schellekens, \np {\bf 289}
(1987) 609; W. Lerche, B.E.W. Nilsson, A.N. Schellekens and N.P. Warner,
\np {\bf 299} (1988) 91; W. Lerche, A.N. Schellekens and N.P. Warner, 
\pr {\bf 177} (1989) 1}
\lref\agnt{I. Antoniadis, E. Gava and K.S. Narain, \np {\bf 383}
(1992) 92; \pl {\bf 283} (1992) 209}

\lref\nsii {H.P. Nilles and S. Stieberger, \np {\bf 499} (1997) 3}
\lref\hm {J.A. Harvey and G. Moore, \np {\bf 463} (1996) 315}
\lref\klm {A. Klemm, W. Lerche and P. Mayr, \pl {\bf 357} (1995) 313}

\lref \DKLI {L. Dixon, V. Kaplunovsky and J. Louis, \np {\bf 329} (1990) 27}
\lref \strominger{A. Strominger, \cmp {\bf 133} (1990) 163}
\lref\cdfp {A. Ceresole, R. d' Auria, S. Ferrara and A. van Proeyen,
\np {\bf 444} (1995) 92}

\lref\all {B. de Wit, P.G. Lauwers, 
R. Philippe, S.Q. Su, A. van Proeyen, \pl {\bf 134} (1984) 37; 
B. de Wit, A. van Proeyen, \np {\bf 245} (1984) 89;
B. de Wit, P.G. Lauwers, A. van Proeyen, \np {\bf 255} (1985) 569; 
E. Cremmer, C. Kounnas, A. van Proeyen, J.P. Derendinger, 
S. Ferrara, B. de Wit, L. Girardello, \np {\bf 250} (1985) 385}

\lref \agntz {I. Antoniadis, E. Gava, K. Narain and T. Taylor,
\np {\bf 432} (1994) 187}
\lref \vk {V. Kaplunovsky, \np {\bf 307} (1988) 145 and hep--th/920570}
\lref \DKLII {L. Dixon, V. Kaplunovsky and J. Louis, \np {\bf 355} (1991) 649}

\lref\dewit {B. de Wit, V. Kaplunovsky, J. Louis and D. L{\"u}st, 
\np {\bf 451} (1995) 53}

\lref\afgnt {I. Antoniadis, S. Ferrara, E. Gava, K. Narain and T. Taylor,
\np {\bf 447} (1995) 35}

\lref\kk{E. Kiritsis and  C. Kounnas, 
\np {\bf 442} (1995) 472; P.M. Petropoulos and J. Rizos, \pl 
{\bf 374} (1996) 49}

\lref\dfkz {J.P. Derendinger, S. Ferrara, C. Kounnas and F. Zwirner, 
\np {\bf 372} (1992) 145;
I. Antoniadis, E. Gava, K.S. Narain and T.R. Taylor,
\np {\bf 407} (1993) 706}

\lref\msi {P. Mayr and S. Stieberger, \np {\bf 407} (1993) 725;
D. Bailin, A. Love, W. Sabra and S. Thomas, 
{Mod. Phys. Lett.} {\bf A9} (1994) 67; {\bf A10} (1995) 337}
\lref\min {J. A. Minahan, \np {\bf 298} (1988) 36; P. Mayr,
S. Stieberger, \np {\bf 412} (1994) 502; K. F{\"o}rger, B.A. Ovrut,
S. Theisen, D. Waldram, \pl {\bf 388} (1996) 512} 

\lref\ejm {J. Ellis, P. Jetzer, L. Mizrachi, \np {\bf 303} (1988) 1}
\lref\gkkopp {A. Gregori, E. Kiritsis, C. Kounnas, N. A. Obers, P. M. 
Petropoulos, B. Pioline, hep--th/9708062}
\lref\vadim{ V. Kaplunovsky, \np {\bf 307} (1988) 145}
\lref\llt{J. Lauer, D. L{\"u}st, S. Theisen, \np {\bf 309} (1988) 771}

\Title{\vbox{\rightline{\tt hep-th/9709004} 
\rightline{NEIP--009/97}    }}
{\vbox{\centerline{String Amplitudes and N=2, $d=4$ Prepotential}
\vskip4pt\centerline{in Heterotic $K3\times T^2$ Compactifications }}}

\centerline{K. F{\"o}rger$^1$\ \ and\ \ S. Stieberger$^2$}

\bigskip\centerline{\it $^1$Sektion Physik}
\centerline{\it Universit{\"a}t M{\"u}nchen}
\centerline{\it Theresienstra\ss e 37}
\centerline{\it D--80333 M{\"u}nchen, FRG}

\vskip .2in
\bigskip\centerline{\it $^2$Institut de Physique Th{\'e}orique}
\centerline{\it Universit{\'e} de Neuch{\^a}tel}
\centerline{\it CH--2000 Neuch{\^a}tel, SWITZERLAND}

\vskip .3in
For the gauge couplings, which arise after toroidal compactification
of six--dimensional heterotic N=1 string theories from the $T^2$ torus,
we calculate their one--loop corrections. This is performed by considering 
string amplitudes involving two gauge fields and moduli fields.
We compare our results with the equations following from N=2 special
geometry and the underlying prepotential of the theory.
Moreover we find relations between derivatives of
the N=2, $d=4$ prepotential and world--sheet $\tau$--integrals which appear
in various string amplitudes of any $T^2$--compactification.
\Date{09/97} 


\newsec{Introduction}
The vector multiplet sector of N=2 supergravity in four dimensions is 
governed by a holomorphic function, namely the
prepotential $F$ \all. 
Up to second order derivatives the Lagrangian is constructed from it.
The K{\"a}hler metric and the
gauge couplings are expressible by derivatives of this prepotential.
Effective N=2 supergravity theories in four dimensions arise e.g. after
compactifications of heterotic string theories on $K3\times T^2$
or type IIA or type IIB on a Calabi--Yau (CY) threefold.
In the following we will concentrate on the first vacuum.
In addition to the heterotic gauge group, which depends on the
specific instanton embeddings and may be --in certain cases-- 
completely Higgsed away,
a $K3\times T^2$ heterotic string compactification always possesses
the $U(1)^2_+\times U(1)^2_-$ internal gauge symmetry. The first factor
comes from the internal graviton and the last factor arises from
the compactification of the six--dimensional tensor multiplet which
describes the heterotic dilaton in six dimensions.
The gauge fields of the left $U(1)_L$'s belong to vector multiplets
whose scalars are the $T$ and $U$ moduli of the torus $T^2$.
Their moduli space is described by special geometry \strominger\DKLI. 
Besides, there is a vector multiplet for the heterotic dilaton and
the graviphoton, whereas the $K3$
moduli and the gauge bundle deformations come in scalars of hyper multiplets.

At special points in the $T,U$ moduli space,
the $U(1)_L^2$ may be enhanced to $SU(2)_L\times U(1)_L$ for $T=U$,
to $SU(2)_L^2$ for $T=U=i$ and $SU(3)_L$ at $T=U=\rho$ where
$\rho=e^{2\pi i/3}$ \higgs.
Logarithmic singularities in the effective string
action appear at these special points. 
This effect should be seen e.g. in the one--loop gauge 
couplings, where heavy strings have been integrated out. Modes, which 
have been integrated out and become
light at these special points in the moduli space are responsible
for these singularities.
These one--loop couplings can be expressed by the perturbative
prepotential.
Therefore, a calculation of one--loop gauge couplings gives
information about the prepotential (and vice versa). Until 
now\foot{Except \msV.}, 
no threshold calculation has been undertaken for gauge groups, where
the modulus and the gauge boson under consideration 
sit in the same vector multiplet.
In the following we focus on the perturbative prepotential
of the vector multiplets.

One--loop gauge threshold corrections are very important quantities for 
at least three aspects of string theory: They play an important
r{\^o}le in constructing consistent, i.e. anomaly free, effective string
actions \jv, they are sensitive quantities for heterotic--typeII string
duality tests \kv\klm\klt\ 
and shed light on the perturbative and hopefully also on the
non--perturbative part of the prepotential \hm.
In this paper we will focus on the last issue: We calculate
threshold corrections to the $U(1)_L^2$ gauge bosons which are
given as certain $\tau$--integrals. The latter we can 
express in terms of the prepotential and its derivatives.
This not only gives a
check of the framework of N=2 supergravity emerging for
heterotic $K3\times T^2$ string compactifications, which tells us, how these
corrections have to be expressed by the prepotential, but it gives
relations between string amplitudes given as world--sheet torus
integrals and the prepotential. Such relations are important for
understanding the structure of string amplitudes in any dimensions
and its relation to a function, which in $d=4$ is the N=2 prepotential.

The organization of the present paper is as follows:
In section 2 we briefly review some facts about N=2 supergravity
in light of the gauge couplings. Section 3 is devoted to a
string derivation of the $U(1)_L^2$--gauge couplings by calculating
the relevant string amplitudes.
In section 4 we find various relations between world--sheet 
$\tau$--integrals and combinations of the prepotential and its 
derivatives. These relations allow us to express the string amplitudes
in terms of the prepotential which is the appropriate form
to compare with the supergravity formulae of section 2.
In section 5 we trace back the origin of the gauge couplings 
to six dimensions via the elliptic genus.
Section 6 gives a summary of our results and some concluding
remarks. All technical details for the $\tau$--integrations are
presented in the appendix.

\newsec{N=2 supergravity and effective string theory}

For N=2,\ $d=4$ heterotic string vacua arising from $K3\times T^2$
compactifications, the dilaton field $S$, the $T$ and $U$ moduli
describing the torus $T^2$ (and possible Wilson lines) are scalar fields
of N=2 vector multiplets \cdfp\dewit. The absence of couplings between 
scalars of vector multiplets and scalars of hyper multiplets 
(describing e.g. the K3 moduli) allows one to study the two moduli
spaces seperately \DKLI.
Since the dilaton field $S$ comes in a vector multiplet, the 
prepotential describing the gauge sector may receive  
space--time perturbative and non--perturbative corrections in
contrast to the hyper multiplet moduli space, which does not get any
perturbative corrections.
However in N=2 supergravity one expects a non--renormalization 
theorem following from chiral superspace integrals which prohibits higher than
one--loop corrections to the prepotential.
Besides, in heterotic string theories, the dilaton obeys a
continuous Peccei--Quinn symmetry to all orders in perturbation 
theory which also forbids higher than one--loop corrections.
Therefore, the only perturbative correction to the tree--level prepotential
comes at one--loop, summarized by a function $f$.  
For the prepotential describing the $S,T$ and $U$
moduli space of a heterotic $K3\times T^2$ compactification one has 
 (neglecting non--perturbative corrections) \dewit\afgnt
\eqn\prep{
F(X)={X^1X^2X^3 \o X^0} +(X^0)^2 f\lf({X^2\o X^0},{X^3\o X^0}\ri)\ ,}
with the unconstrained vector multiplets $X^I,\ I=0,\ldots,3$. 
The function $f$ is much constrained by the perturbative duality 
group $SL(2,\IZ)_T\times
SL(2,\IZ)_U\times \IZ^{T\leftrightarrow U}_2$. The field
$X^0$ accounts for the additional vector multiplet of
the graviphoton. The gauge kinetic part of the Lagrangian is
\eqn\egaugea{
\Lc_{\rm gauge}=-{i\o 4}(\Nc_{IJ}-\bar\Nc_{IJ}) F_{\mu\nu}^I F^{\mu\nu\,J}+
{1\o 4}(\Nc_{IJ}+\bar\Nc_{IJ}) F_{\mu\nu}^I \tilde F^{\mu\nu\,J}\ ,}
where the gauge couplings are then expressed in terms of ${\cal N}_{IJ}$
\eqn\proyen{
{\cal N}_{IJ}=\bar F_{IJ}+2i {{\im F_{IK} \im F_{JL} X^K X^L} \o {\im F_{MN}
X^M X^N}}  }
via
\eqn\gauge{
g^{-2}_{IJ}={\cal N}_{IJ}-{\ov {\cal N}}_{IJ}\ .}
The scalar partner (and spinor) of the graviphoton is gauge fixed 
in super Poincar{\'e} gravity. 
With the standard choice for the scalar fields of the vector multiplets
\eqn\moduli{
S={X^1 \o X^0}\ \ \ ,\ \ \ T={X^2 \o X^0}\ \ \ ,\ \ \
U={X^3 \o X^0}\ ,}
we derive from \proyen\ and \gauge\ for the effective gauge couplings
up to order $\Oc(f^2,(\p f)^2/(S-\ov S)$
\eqn\gauges{\eqalign{
g^{-2}_{22}&={(S-\bar S)(U-\bar U)\o (T-\bar
 T)}-{1\o 4}\Big[\p_TD_Tf-\p_{\bar T}D_{\bar T}\bar f\Big]-
{1\o 4}{(U-\bar U)^2\o(T-\bar T)^2}\Big[\p_U D_U f-\p_{\bar U}D_{\bar
 U}\bar f\Big]\cr
&+\h{(U-\bar U)\o(T-\bar T)}\Big[\p_T\p_U f-\p_{\bar T}
\p_{\bar U}\bar f\Big]+\Oc\lf(f^2,(\p f)^2\o S-\ov S\ri) \cr
g^{-2}_{23}&=(S-\bar S)-\h\Big[D_TD_Uf-D_{\bar T}D_{\bar U}
\bar f\Big]
+{1\o 4}{(T-\bar T)\o(U-\bar U)}\Big[\p_T D_T f-\p_{\bar T} D_{\bar T}
\bar f \Big]\cr
&+{1\o 4}{(U-\bar U)\o(T-\bar T)}\Big[\p_UD_U f-\p_{\bar U}D_{\bar
 U}\bar f\Big]+\Oc\lf(f^2,(\p f)^2\o S-\ov S\ri) \ ,}}
where the covariant derivatives are
$D_T=\p_T-{2\o (T-\bar T)}$ and $D_{\bar T}=\p_{\bar T}+{2\o (T-\bar
T)}$
and the indices $T,U$ correspond to $I=2,3$ respectively.
Of course, there is no dilaton dependence at one--loop. 
The expression for $g^{-2}_{33}$ we simply deduce from 
$g^{-2}_{22}$ by exchanging $T$ and $U$.
The couplings in \gauges\ are related via 
${(T-\bar T)\o (U-\bar U)}g^{-2}_{22}+g_{23}^{-2}=
2(S-\bar S)-\h\Big(D_TD_Uf-\p_T\p_U f-hc.\Big)$.

In the interpretation of \gauges\ one occurs a puzzle:
Along the explanations of above the gauge--couplings should not
receive higher than one--loop contributions, i.e. 
powers in $1/(S -\ov S)$ must not appear.
Nonetheless, it is the dilaton independent part of \gauges, which is
relevant for the string one--loop calculations in the next section.
This is in precise analogy of \afgnt, where an expansion like
in \gauges\ has been performed for the one--loop K{\"a}hler metric.
All the same, for completness, let us mention the solution to that puzzle 
in view of the gauge--couplings \proyen. The symplectic 
transformation $(X^I,F_J)\rightarrow(\hat{X}^I,\hat{F}_J)$ \dewit
\eqn\symplectic{\eqalign{
\hat{X}^1=F_1&,\ \ \hat{F}_1=-X^1\ ,\cr
\hat{X}^I=X^I&,\ \ \hat{F}_I=F_I\ , I\neq 1}}
changes the metric from ${\cal N}$ to 
$\hat{\cal N}$ with \def\n {{\cal N}}
\eqn\transmat{
\hat{\cal N}=\pmatrix{\n_{00}-{\n_{01}\n_{10}\o
\n_{11}}&{\n_{01}\o\n_{11}}&\n_{02}-{\n_{01}\n_{12}\o\n_{11}}&\n_{03}-
{\n_{01}\n_{13}\o\n_{11}}\cr\noalign{\vskip2pt}
{\n_{10}\o\n_{11}}&-{1\o\n_{11}}&{\n_{12}\o\n_{11}}&{\n_{13}\o\n_{11}}
\cr\noalign{\vskip2pt}
\n_{20}-{\n_{10}\n_{21}\o
\n_{11}}&{\n_{21}\o\n_{11}}&\n_{22}-{\n_{12}\n_{21}\o\n_{11}}&\n_{23}-
{\n_{13}\n_{21}\o\n_{11}}\cr\noalign{\vskip2pt}
\n_{30}-{\n_{10}\n_{31}\o
\n_{11}}&{\n_{31}\o\n_{11}}&\n_{32}-{\n_{12}\n_{31}\o\n_{11}}&\n_{34}-
{\n_{13}\n_{31}\o\n_{11}}\cr}\ .}
It can be verified that in this new basis \transmat, all gauge
couplings involve neither powers of $1/(S-\ov S)$ nor higher orders in
$f$ or its derivatives. This is just an effect of a rearrangement of all
couplings $\n_{IJ}$ \proyen\ in \transmat.
E.g. for ${\hat{g}^{-2}}_{22}\equiv{\hat{\n}}_{22}-
{\ov{\hat{\n}}_{22}}=
2i\im\lf(\n_{22}-{\n_{12}^2\o\n_{11}}\ri)$ one determines
\eqn\newgTT{\eqalign{
{\hat{g}^{-2}}_{22}
&=-4(\tilde S-\ov{\tilde S}) {|U|^2\o (T-\ov T)(U-\ov U)}\cr
&-{1\o(U-\ov U)^2}\lf[
\ov U^2\p_T D_T f-U^2\p_{\ov T} D_{\ov T} \ov f\ri]-{1\o(T-\ov T)^2}
\lf[U^2 \p_UD_U f-\ov U^2\p_{\ov U} D_{\ov U} \ov f\ri]\ ,\cr
{\hat{g}^{-2}}_{23}&=-(\tilde S-\ov{\tilde S}) 
{(T+\ov T)(U+\ov U)\o (T-\ov T)(U-\ov U)}-\h\lf[D_UD_Tf-D_{\ov
U}D_{\ov T}\ov f\ri]\cr
&-{1\o(U-\ov U)^2}\lf[
T\ov U\p_T D_T f-\ov T U\p_{\ov T} D_{\ov T} \ov f\ri]-
{1\o(T-\ov T)^2}\lf[\ov T U \p_U D_U f-T\ov U \p_{\ov U} D_{\ov U}\ov
f \ri]\ ,}}
with the pseudo--invariant dilaton $\tilde S$ \dewit
\eqn\psS{
\tilde S=S+\h \p_T\p_U f\ .}

\newsec{String amplitudes}
In this section we determine the one--loop correction
to the $U(1)^2_L$ gauge couplings of the effective action of
an N=2 heterotic string compactified on $K3\times T^2$ which has 
been studied in \dewit\ from a field theoretical point of view.
Here we want to focus on the derivation via string amplitudes.
To this end we calculate the CP even part of
two--point one--loop string amplitudes including gauge bosons 
of the internal Abelian gauge group of the torus $T^2$ in a background
field method.
Then we compare the $\Oc(k^2)$ piece of the string amplitudes with
the effective Lagrangian \egaugea\  of N=2 supergravity. 
We also compute three point amplitudes with two gauge bosons
and a modulus $U$ or $T$ which corresponds to derivatives of
the gauge couplings.

The relevant vertex operators in the zero ghost picture
for the moduli 
$T=T_1+iT_2=2(b+i\sqrt{G})$ and $U=(G_{12}+i\sqrt{G})/G_{11}$
w.r.t. the
background fields $G_{IJ}$ and $B_{IJ}=b\, \eps_{IJ}$ are
\eqn\vertm {
V_\pm^{(0)}= \bar\p X^\pm\,\Big[\p X^\pm+i (k\cdot \psi)\Psi^\pm\Big] 
e^{i k\cdot X}\ ,}
where $X^\pm={1\o \sqrt{2}}\Big(X_1\pm iX_2\Big)$ 
are the internal bosonic fields, $\Psi^\pm$
their supersymmetric partners and $\psi^\mu$ are spacetime fermions
with $\mu=0,\cdots,3$.
The vertex operators for the $U(1)_L^2$ gauge bosons of the internal torus
are \llt
\eqn\vertg{
V_{A_\pm}^{(0)}=\rho\ \eps_\mu\bar\p X^\pm\Big[\p X^\mu+i (k\cdot 
\psi)\psi^\mu\Big]
 e^{i k\cdot X}\ ,}
with spacetime polarization tensor $\eps_\mu$ and 
$\rho(T)=\sqrt{U_2\o T_2}$ and $\rho(U)=\sqrt{T_2\o U_2}$.

There is an important point regarding the choice of 
normalization of the vertex operators, which has two, seemingly
different, explanations: one
based on target--space--duality, i.e. string theory and one
coming from N=2 supergravity.
The stringy argument: The calculated amplitudes
have to have a certain modular weight under $T$-- and $U$--duality as
it can be anticipated from \newgTT. This
is precisely achieved by that choice.
The supergravity argument: The specific mixing between the
scalars of the vector multiplet and gauge bosons via the covariant derivative
involves a coupling which is not the gauge coupling but given by the 
K{\"a}hler metric.
If the fields and propagators are correctly normalized the 
corresponding Feynman diagram contributes only with the 
gauge coupling.

\subsec{Two-point string amplitudes}
We consider the $\Oc(k^2)$ contribution of 
the two point one--loop string amplitude  
including two gauge bosons $A_\mu^+$. It will produce a term
\eqn\stgauge
{-{i\o 4}{\Delta_{(TT)}\o 8\pi^2}F_{\mu\nu}^TF^{\mu\nu\,T}} 
in the effective action. We
denote the one--loop threshold correction to the internal $U(1)_T$ 
gauge coupling
by $\Delta_{(TT)}$. On the other hand, the 
gauge couplings \gauges, which refer to the supergravity basis, will
turn out to be linear combinations of the couplings of 
$U(1)_T$ and $U(1)_U$.

We take the gauge boson vertex operators of the two--point function
in a constant background field similarly to \vadim. 
Otherwise,  the kinematic factor  will cause the two point amplitude
to vanish. 
Thus we take 
$A_\mu^+(X)=-\h F_{\mu\nu}^T X^\nu$ with $F_{\mu\nu}^T={\rm const}$ and
the polarization tensor of the gauge boson $A_\mu^+$ is 
replaced by  $\eps_\mu e^{i k\cdot X}\to A^+_\mu(X)$. The vertex
operator of the gauge bosons is then
$\tilde V^{(0)}_{A_+}=-\h F_{\mu\nu}^T\rho(T)\bar\p X^+(X^\nu \p X^\mu+
\psi^\mu\psi^\nu)$.
The general  expression for the CP even part of the
string amplitude is \DKLII 
\eqn\eampl{\Ac(A_+,A_+)=\sum_{\rm even\, s}(-1)^{s_1+s_2}\int_{\Gamma} d^2\tau
\,Z(\tau,\bar\tau,s)\int d^2z_1\la
\tilde V_{A_+}^{(0)}(z,\bar z)\tilde V_{A_+}^{(0)}(0)\ra\ ,}
where 
\eqn\part{
Z(\tau,\bar\tau,s)={\rm Tr}_{s_1}\lf[(-1)^{s_2 F} q^{H-\h}\bar
q^{\bar H-1}\ri]=Z_\psi Z_X Z_{X_0} Z_{\rm int} }
is the partition function ($q=e^{2\pi i\tau},\ov q=e^{-2\pi i\ov\tau}$)
for even spin structures $(s_1,s_2)=(1,0),(0,0),(0,1)$ and
$Z_\psi={\th_\alpha(0,\tau)\o \eta(\tau)}$ 
the fermionic partition function  where $\th_\alpha$
are the Riemann theta functions for $\alpha=2,3,4$.
The contribution
from bosonic zero modes is $Z_{X_0}={1\o 32 \pi^4 \tau_2^2}$
and $Z_X={1\o |\eta(\tau)|^4}$ is the bosonic partition function.
The fermion number is denoted by $F$.
The integration region is the fundamental region of the worldsheet torus
$\Gamma=\{\tau :\ |\tau_1|\le\h,|\tau|\ge 1\}$.

After summing over even spin structures
we only get non vanishing contributions 
if four space-time fermions are contracted because of 
a theta function identity. Therefore, pure bosonic contractions may be omitted.
The two point function gives\foot{We
introduced the notation $X_i=X(z_i,\bar z_i)$.}
\eqn\tt
{\la \tilde V_{A_+}^{(0)}(\bar z,z)\tilde V_{A_+}^{(0)}(0)\ra=-\h
F_{\mu\nu}^TF^{\mu\nu\,T}\rho(T)^2 G_F^2\la\bar \p X_1^+\bar
\p X^+_2\ra\ ,}
where 
$G_F={\th_1(0,\tau)\th_\alpha(z,\tau)\o \th_\alpha(0,\tau)\th_1(z,\tau)}$
with $\alpha=2,3,4$ is
the fermionic Green function and the part of
$G_F^2$ depending on spin structures is $4\pi i\p_{\tau}\ln Z_\psi$
which does no longer depend on worldsheet coordinates.
$G_B=-\ln|\chi|^2$ is the
bosonic Green function with $|\chi|^2=4\pi^2 e^{-2\pi (\im
\,z)^2/\im\tau}\Big|{\th_1(z,\tau)\o\th_1(0,\tau)}\Big|^2$. 
We take the following Green functions for the internal bosons 
\eqn\ig{
\eqalign{
\la\bar\p X^\pm\bar\p X^\pm\ra&=2\pi^2(P_R^\pm)^2\cr
\la\bar\p X^\pm\bar\p X^\mp\ra &=2\pi^2 P_R^\pm P_R^\mp-{\pi\o\tau_2}
+\bar \p^2 G_B\ ,}}
with Narain momenta $P_{R/L}^+=\bar P_{R/L}$ and $P_{R/L}^-=P_{R/L}$
which are defined as
\eqn\momenta{
\eqalign{
P_L&={1\o \sqrt{2T_2 U_2}}\Big(m_1+m_2\bar U+n_1\bar T+n_2\bar T\bar U\Big)\cr
P_R&={1\o \sqrt{2T_2 U_2}}\Big(m_1+m_2\bar U+n_1 T+n_2 T\bar U\Big)\ .}}
From $\int d^2z \,\p^2 G_B=0$ we get
$\int d^2 z\, \p^2 \ln\th_1(z,\tau)=-\pi$. Using this relation
we find the following result
\eqn\att{
\Ac(A_+,A_+)=-{i\o 4} F_{\mu\nu}^TF^{\mu\nu\,T}\Delta_{(TT)}\ ,}
with
\eqn\dtt{
\Delta_{(TT)}=-{U_2\o T_2}\int 
{d^2\tau\o\tau_2} 
\bigg[\sum_{P_L,P_R} q^{\h |P_L|^2}\bar q^{\h |P_R|^2}
\bar P_R^2\bigg]\bar F_{-2}(\bar\tau)\ ,}
where $\ov F_{-2}={\ov E_4\ov E_6\o \ov\eta^{24}}$ and
we define
\eqn\torus{
Z_{torus}(\tau,\ov\tau)=\sum_{(P_L,P_R)}q^{\h |P_L|^2}
\ov q^{\h |P_R|^2}:=\sum_{(P_L,P_R)} \hat{Z}_{torus}\ .}

The string amplitude involving
$\la \tilde V^{(0)}_{A_-}\tilde V^{(0)}_{A_-}\ra$ is easily 
obtained from \att\ 
by exchanging $T_2$ with $U_2$ and replacing $\bar P_R$ with its
complex conjugate $P_R$.
Similarly, for the string amplitude $\la
\tilde V^{(0)}_{A_+} \tilde V^{(0)}_{A_-}\ra$ we get the modular invariant 
result: 
\eqn\atu{
\Ac(A_+,A_-)={i\o 4} F_{\mu\nu}^TF^{\mu\nu\,U}\int{d^2\tau\o
\tau_2}\sum_{(P_L,P_R)}
\Big(|P_R|^2-{1\o 2\pi\tau_2}\Big)
\hat{Z}_{torus}\bar F_{-2}(\bar\tau)\ .}
This result can be directly compared with the one loop correction to the
K{\"a}hler potential 
$G_{T\bar T}^{(1)}=-{i\o 2}G_{T\bar T}^{(0)} D_T D_U f+hc.$ which has been
derived in \afgnt.
The second part may be identified with the Green-Schwarz (GS)--term 
$2 G^{(1)}\equiv\Delta_{\rm univ}=\int {d^2\tau\o \tau_2} \Big(-{1\o
2\pi\tau_2}\Big) Z_{torus}\bar F_{-2}$.

Our results \att\ and \atu\ refer to the string basis (3.1) and (3.2) and
(therefore) involve modular invariant integrands.
Since the momenta transform under $SL(2,{\bf Z})_T\times SL(2,{\bf
Z})_U$, like $(ad-bc=1)$
\eqn\dulnar{\eqalign{
(P_L,\bar P_R)&\to\sqrt{cT+d \o c\bar T+d}\ (P_L,\bar P_R)\ \ \ ,\ \ \
T\to {{aT+b}\o {c T+d}}\ , \ U\to U\ , \cr
(P_L,P_R)&\to\sqrt{cU+d\o c\bar U+d}\ (P_L,P_R)\ \ \ ,\ \ \ 
T\to T\ ,\ U\to {{a U+b} \o {c U+d}\ }\ ,}}
we realize that these amplitudes transform with specific weights
$(w_T,w_U)=(2,-2)$, $(w_{\bar T},w_{\bar U})=(0,0)$ and 
$(w_T,w_U)=(0,0)$,  $(w_{\bar T},w_{\bar U})=(0,0)$, respectively.
They can be directly
identified with well--defined integrals $\Ic_{2,-2}$ and $\Ic_{0,0}$
as will be shown in the next section. 

The one--loop correction to the gauge coupling $g_{22}^{-2}$, as it
has appeared in the last section and which is therefore w.r.t. 
the supergravity basis, is then obtained 
by taking a linear combination of string amplitudes \att\ and \atu,
which corresponds
to the correlation function of two  gauge boson vertex operators  
$A_\mu^{T}=\h[A_{\mu}^+-{(U-\bar U)\o (T-\bar T)}A_\mu^-]$:
\eqn\gloop
{\eqalign{
\Big[g_{22}^{-2}\Big]^{1-loop}&={1\o 4}
\bigg[{\Delta_{(TT)}\o 8\pi^2}-
2{(U-\bar U)\o(T-\bar T)}{\Delta_{(TU)}\o 8\pi^2}
+{(U-\bar U)^2\o (T-\bar T)^2}
{\Delta_{(UU)}\o 8\pi^2}\bigg]-{1\o 8\pi^2}{U_2\o T_2}G^{(1)}\cr
&=-{1\o 32\pi^2}\bigg[{U-\bar U\o T-\bar T}\int {d^2\tau\o\tau_2}
\sum_{(P_L,P_R)} (\bar P_R-P_R)^2 \hat Z_{torus}\bar
F_{-2}(\bar\tau)\bigg]\ .}}
Notify, that in the above expression 
the GS-term cancels the one in $\Delta_{(TU)}$.
After symplectic transformation \symplectic\ to the gauge coupling 
${\hat g}_{22}^{-2}$ one obtains for its loop--correction:
\eqn\ghloop
{\eqalign
{\Big[{\hat{g}}_{TT}^{-2}\Big]^{1-loop}&=
{{\bar U}^2\o (U-\bar U)^2}{\Delta_{(TT)}\o 8\pi^2}
+{U^2\o (T-\bar T)^2}
{\Delta_{(UU)}\o 8\pi^2}+2{|U|^2\o (U-\bar U)(T-\bar T)}
{\Delta_{(TU)}\o 8\pi^2}\cr
&+{|U|^2\o (T-\bar T)(U-\bar U)}
{G^{(1)}\o 2\pi^2}\cr
&=-{1\o 8\pi^2}\bigg[{1\o (T-\bar T)(U-\bar U)}\int {d^2\tau\o\tau_2}
\sum_{(P_L,P_R)} (\bar U\bar P_R+U P_R)^2 \hat Z_{torus}\bar
F_{-2}(\bar\tau)\bigg]\ .}}
This amplitude can be directly derived from a two point amplitude with
vertex operators corresponding to 
$\hat{A}_\mu^T={\bar U\o(U-\bar U)}A_\mu^+-{U\o (T-\bar T)}A_\mu^-$.
Using the results from the next section we may directly cast
\gloop\ and \ghloop\ into the forms (2.6) and (2.9), dictated by
supergravity.

In the corrections \gloop\ and \ghloop, there appears the non--modular
invariant GS--term $G^{(1)}$. On the other hand, the tree--level  
dilaton field gets modified at one--loop by the same amount with an 
opposite sign \dfkz. It is the one--loop dilaton $S$ which appears
in (2.9):

\eqn\dila{
g_{\rm string}^{-2}={S-\ov S\o 2}-{1\o 16\pi^2}G^{(1)}\ .}
Thus, altogether, the physical coupling stays modular invariant.
See also \nsii\ for a more complete discussion. 

\subsec{Three-point amplitudes}
Now we consider three point amplitudes which involve two internal
gauge bosons and one modulus. 
First we want to investigate the amplitude including two gauge bosons
$A_T$ and one $T$ modulus 
which is related to $ \p_T\Delta_{(TT)} F_{\mu\nu}F^{\mu\nu}T$ in the
effective string action, where $\p_T\Delta_{(TT)}$ denotes
the derivative with respect to $T$ of the one loop correction to the 
$U(1)_T$ gauge coupling. 
The correlation function gives the following contractions:
\eqn\ttt
{\eqalign{\la V_+^{(0)}(z_1)V_{A_+}^{(0)}(z_2)V_{A_+}^{(0)}(z_3)\ra&=\Kc
\,G_F^2\prod_{i<j}|\chi_{ij}|^{2 k_i\cdot k_j} {U_2\o i T_2^2} \cr
&\Big(\la\bar\p X_1^+\p X_1^+\ra\la\bar \p X_2^+\bar \p X_3^+\ra
+\la\bar\p X_1^+\bar \p X_2^+\ra\la\p X_1^+\bar \p X_3^+\ra\cr
&+\la\bar\p X_1^+\bar
\p X_3^+\ra\la\p X_1^+\bar\p X_2^+\ra\Big) ,}}
where $\Kc=\Big((k_2k_3)(\eps_2\eps_3)-(k_2\eps_3)(k_3\eps_2)\Big)$ is
the kinematic factor.

Before doing the worldsheet integrals we want to make some comments
on possible additional non trivial contributions to the $\Oc(k^2)$ part of
the amplitude. 
We may get contributions from the
delta function which might appear in the correlation function $\la \bar
\p X^\pm\p X^\mp\ra$. If we consider the region
$|z_{ij}|<\eps$ then $|\chi_{ij}|\simeq |z_{ij}|$ and thus the delta function
can be omitted because  $\int d^2z_i\,
\delta^{(2)}(z_{ij}) |z_{ij}|^{2 k_i\cdot k_j} f(z_{ik})=0$ where
$f$ is some function. 
But if $|z_{ij}|> \eps$ and $|k_i\cdot k_j G_{ij}^B|\ll 1$ then
one can expand $|\chi_{ij}|^{2 k_i\cdot k_j}=1-k_i\cdot k_j
G_{ij}^B+\ldots$  and in this case one indeed gets contributions
from the delta function for the lowest term of the expansion
\ejm.
On the other hand, if the correlation functions can be approximated such
that the worldsheet integral gives
$\int_{|z_{il}|<\eps} d^2 z_{il} {|z_{il}|^{2
k_i\cdot k_l}\o |z_{il}|^2}\sim {\pi\o k_i\cdot k_l}$ 
one may e.g. produce a $\Oc(k^2)$ contribution from  terms of the
order $\Oc(k^4)$. These
contributions are important when one has to collect 
all possible terms of a particular order \min.
But in the case considered here,  pinched off integrals 
only give $\int_{|z_{il}|<\eps} d^2 z_{il} {|z_{il}|^{2
k_i\cdot k_l}\o |z_{il}|^4}\simeq  {\pi\o k_i\cdot k_l-1}$ and thus do
not contribute to the $\Oc(k^2)$ piece of the amplitude. 

In the following we will restrict ourselves to the region
$|z_{ij}|<\eps$. Taking into account the arguments mentioned above 
it remains to perform the worldsheet integral of  \ttt. 
We end up with:
\eqn\attt{
\Ac(T,A_+,A_+)|_{\Oc(k^2)}=-\Kc{\pi^2 \ U_2\o 2\  T_2^2}\int
d^2\tau
\bigg\{\sum_{P_L,P_R} q^{\h |P_L|^2}\bar q^{\h |P_R|^2} \bar P_R^3
P_L \bigg\} 
\bar F_{-2}(\bar\tau)\,}
This term can be identified with the third
derivative of the prepotential\foot{The relevant relations between the 
prepotential and $\tau$--integrals may be found in the next section.} 
$f_{TTT}$ which  has been derived in \afgnt \ by taking particular
derivatives on the integral coming from a CP odd  string amplitude
of the one loop correction to the K{\"a}hler potential
$G_{T\bar T}^{(1)}$. 
Thus one finds
\eqn\fattt{
\Ac(T,A_T,A_T)|_{\Oc(k^2)}=-4 i\Kc\pi^3 f_{TTT}\ .}
We will have to say more about this result in section 4.
We realize that this expression transforms covariantly 
under $SL(2,{\bf Z})_T\times SL(2,{\bf Z})_U$
with weights $(w_T=4,w_{\bar T}=0)$
and $(w_U=-2,w_{\bar U}=0)$, respectively.

Besides we calculate the three point amplitudes 
$\la V_+^{(0)}V^{(0)}_{A_+}V_{A_-}^{(0)}\ra$ and
$\la V_+^{(0)}V^{(0)}_{A_-}V_{A_-}^{(0)}\ra$
with the result
\eqn\attu
{\eqalign
{\Ac(TA_+ A_-)&=-{\Kc\pi^2\o 2\ T_2}\int d^2\tau
\bigg\{\sum_{P_L,P_R} q^{\h |P_L|^2}\bar q^{\h |P_R|^2}
\Big[ \bar P_R |P_R|^2 P_L-{1 \o \pi\tau_2}\bar P_R P_L\Big]\bigg\} 
\bar F_{-2}(\bar\tau)\cr
\Ac(TA_- A_-)&=-{\Kc\pi^2\o 2\ U_2}
\int d^2\tau
\bigg\{\sum_{P_L,P_R} q^{\h |P_L|^2}\bar q^{\h |P_R|^2}
\Big[P_R |P_R|^2 P_L-{1\o \pi\tau_2}P_RP_L
\Big]\bigg\} \bar F_{-2}(\bar\tau)\ .}}
These amplitudes can be casted into the convenient form:
\eqn\ampli{\eqalign{
\Ac(TA_+A_-)&=-4i\Kc\pi^3
\bigg[ f_{TTU}+ {1\o 4\pi^2}G_T^{(1)}\bigg]\cr
\Ac(TA_-A_-)&=-4i\Kc\pi^3
\bigg[ f_{TUU}+ {1\o 4\pi^2}G_U^{(1)}\bigg]\ .}}

The linear combination which corresponds to the three point amplitude
with the $T$ modulus  and two $A_\mu^{T}$ gauge bosons is
\eqn\lincomb{\kern-1em\eqalign{
\Ac(T A^{T}A^{T})&={1\o 4}\bigg[\Ac(TA_+A_+)-2{(U-\bar U)\o (T-\bar T)}
\Ac(TA_+A_-)+{(U-\bar U)^2\o (T-\bar T)^2}\Ac(TA_-A_-)\bigg]\cr
&=-{\Kc\pi^2\o 8}{U_2\o T_2^2}\int d^2\tau
\bigg\{\sum_{P_L,P_R} q^{\h |P_L|^2}\bar q^{\h |P_R|^2}
P_L \bar P_R(\bar P_R-P_R)^2\cr
&+{2\o \pi\tau_2}P_L\bar P_R-{1\o\pi\tau_2}P_LP_R
\bigg\} \bar F_{-2}(\bar\tau)\ .}}
The string amplitude is not a 1PI diagram but 
also contains other exchange diagrams and 
therefore splits into field theoretical amplitudes containing 
one loop corrections to the gauge coupling.
\eqn\ft{\kern-1em
\Ac(T A^{T}A^{T})={\Kc\pi U_2\o 4 i T_2}\lf[16\pi^2\p_T g_{TU}^{-2}
 -{1\o 4 i U_2}\Delta_{(TT)}+{U_2\o 4 i T_2^2}\Delta_{(UU)}\ri]\ .}

\newsec{Prepotential and world--sheet torus integrations}

In this section we want to find relations of the one--loop
prepotential $f$ and/or derivatives of it to world--sheet
$\tau$--integrals as they appear in the previous section.
The one--loop correction\foot{Compared to the 
previous sections
we now change $f\to i f$.} to the heterotic prepotential \prep\
can be written in the chamber $T_2>U_2$ \hm

\eqn\fone{
f(T,U)={i\o 6\pi }U^3
+{2\o (2\pi i)^4}\sum_{(k,l)>0} c_1(kl)\ \Li_3 \lf[e^{2\pi
i(kT+lU)}\ri] 
+{1\o (2\pi)^4} c_1(0)\zeta(3)\ ,}
where the numbers $c_1(n)$ are related to the (new) supersymmetric index 

\eqn\susyindex{
{\cal Z}(\tau,\ov\tau)=\ov \eta^{-2}\ {\rm Tr}_R\lf[F(-1)^F\ 
q^{L_0-{c\o 24}}\ov q^{\ov L_0-{\ov c\o 24}}\ri]\ ,}
which for heterotic compactifications on $K3\times T^2$ with 
the choice of $SU(2)$ instanton numbers 
$(12,12),\ (11,12)\ ,(10,14)$ and $(24,0)$ becomes \hm
\eqn\index{\eqalign{
{\cal Z}(\tau,\ov\tau)&=2i\ Z_{torus}(\tau,\ov\tau)
{\bar E_4 \bar E_6 \o \ov \eta^{24}}\ ,\cr
{\bar E_4 \bar E_6 \o \ov \eta^{24}}&=\sum_{n\geq -1} c_1(n)\ \bar
q^n\ .}}
The mentioned models lead to the
gauge group $E_7\times E_7$ and $E_7\times E_8$, respectively.
In the first three cases the gauge group may be completely Higgsed away.
At the perturbative level these models are equivalent. A fact, which
also becomes clear from the unique expression for the supersymmetric 
index \index\ which enters all kinds of perturbative string
calculations (cf. e.g. the previous section). These three models
(after Higgsing completely) are 
dual to typeIIA Calabi--Yau compactifications, which are elliptic 
fibrations over the Hirzebruch surfaces $\IF_0,\IF_1,\IF_2$. 
Then the holomorphic part (to be identified with the Wilsonian
coupling) of the three-point functions \attt\ and \attu\ 
[in particular (3.18) and (3.20)] is related to the Yukawa couplings 
$f_{TTT}, f_{TTU}$ and $f_{TUU}$ 
of the Calabi--Yau manifold, respectively \klm\klt. 
Moreover using mirror symmetry
these couplings are given by the classical intersection numbers of the
typeIIB theory.
Supersymmetric indices \susyindex, valid for the other
bases $\IF_k$ are the subject of \st. They allow for more general
instanton embeddings and one ends up with larger terminal gauge groups after
Higgsing.

We introduce the polylogarithms $(a \geq 1)$:
\eqn\LIS{
\Li_a(x)=\sum_{p>0} {x^p \o p^a}\ .}

The integrals we should look for involve Narain momenta from `charge' 
insertions or zero--mode contributions.
In general they show up
in string amplitudes involving $U(1)$--charges w.r.t. the internal 
bosonic fields or after contractions of bosonic internal 
fields (belonging to the $T^2$).
I.e. we consider  ($\alpha,\beta,\gamma,\delta \geq 0$)
\eqn\setup{
{\cal I}_{w_T,w_U}:=
(T- \ov T)^m (U-\ov U)^n\ \int {d^2\tau \o \tau_2^k}\  \sum_{(P_L,P_R)}
P_L^\alpha P_R^\beta \ov P_L^\gamma \ov P_R^\delta\   
{\hat Z}_{torus}(\tau,\ov\tau)  \ \ov F_{l}(\ov \tau)\ .} 
We want this expression to have modular weights $(w_T,w_U)$
and weights $(w_{\ov T},w_{\ov U})=(0,0)$ under $T,U$--duality.
This imposes the conditions [cf. \dulnar]:

\eqn\conds{\eqalign{
m&=-{w_T \o 2}\cr
n&=-{w_U \o 2}\cr
\gamma&=\alpha-\h(w_T+w_U)\cr
\delta&=\beta +\h(w_T-w_U)\ .}}
There is also a relation for $k$ and $l$ which follows from modular
invariance of the integrand, which can be easily deduced after a
Poisson resummation on the momenta $m_i$.
Since the integrals \setup\ will be constructed such that they 
transform with a certain weight under $SL(2,\IZ)_T\times SL(2,\IZ)_U$
we expect that ${\cal I}_{w_T,w_U}$ can be written in terms of modular
covariant derivatives of $f$ rather than usual derivatives.
The prepotential $f(T,U)$ has weights $(w_T,w_U)=(-2,-2)$.
Acting with the covariant derivative (cf. also section 2)
\eqn\covT{
D_T=\p_T-{2\o T-\ov T}}
increases its weight $w_T$ by $2$.
In general with the derivative
\eqn\covsT{
D^n_T=\p_T-{2n\o T-\ov T}}
one changes the weight from $-2n$ to $-2n+2$.
This derivative is also covariant w.r.t.
the K{\"a}hler connection $a_\mu \sim [\p_i K(\Phi,\bar\Phi)D_\mu\phi^i-
\p_{\bar i} K(\Phi,\bar\Phi)D_\mu \bar \phi^{\bar i}]$, which means
from the point of view of amplitudes that one--particle reducible
diagrams with massless states running in the loop are subtracted 
to end up with the 1PI effective action.

In subsection 4.1 we consider cases involving only two momenta, i.e.
$\alpha+\beta+\gamma+\delta=2$. 
In subsection 4.2. some cases of more than two momenta insertions.

\subsec{Two momenta insertions}
\subsubsec{4.1.1.\  $f_{TT}$}

Let us consider the integral

\eqn\ftt{
{\cal I}_{2,-2}:=
{(U- \ov U)\o  (T-\ov T)}\ \int {d^2\tau \o \tau_2}\  \sum_{(P_L,P_R)}
\ov P_R^2\   
{\hat Z}_{torus}(\tau,\ov\tau)\ {\ov F}_{-2}(\ov \tau)\ ,} 
which after a Poisson resummation on $m_i$ (cf. Appendix A) becomes

\eqn\ftts{
{\cal I}_{2,-2}=
{(U- \ov U)\o  (T-\ov T)}\ T_2^2
\int {d^2\tau \o \tau_2^4}\ \sum_{n_1,n_2 \atop  l_1,l_2}
\ov Q_R^2\  e^{-2\pi i \ov T \det A}  e^{{-{\pi T_2} \o {\tau_2
U_2}}\lf|n_1\tau+n_2U\tau -Ul_1+l_2 \ri|^2} 
 \ {\ov F}_{-2}(\ov \tau)\ .} 
with $A=\lf( {n_1 \atop n_2}{-l_2\atop l_1}\ri)\in M(2,\IZ)$ and we introduce:
\eqn\Qs{\eqalign{
Q_R&={1 \o \sqrt{2T_2U_2}}\lf[(n_2 \ov U+n_1)\tau-\ov U l_1+l_2\ri] \cr
\ov Q_R&={1 \o \sqrt{2T_2U_2}}\lf[(n_2 U+n_1)\tau-U l_1+l_2\ri] \cr
Q_L&={1 \o \sqrt{2T_2U_2}}\lf[(n_2 \ov U +n_1)\ov\tau-\ov U l_1+l_2\ri]\cr
\ov Q_L&={1 \o \sqrt{2T_2U_2}}\lf[(n_2 U+n_1)\ov\tau-U l_1+l_2\ri]\ .}} 
In that form \ftts\ one easily checks modular invariance, i.e.
one deduces the only possible choice for $k$ and $l$ in \setup.
For the anti--holomorphic function we choose
\eqn\Fanti{
\ov F_{-2}(\ov \tau)={\bar E_4 \bar E_6 \o \bar\eta^{24}}} 
as it arises in physical amplitudes (cf. e.g. section 3).
Modular invariance also enables us to use the orbit decomposition 
used in \DKLII, i.e. decomposing the set of all
matrices $A$ into orbits of $SL(2,\IZ)$:
\eqn\orbits{\eqalign{
I_0&:\ A=\lf( {0 \atop 0}\ {0\atop 0}\ri)\ , \cr
I_1&:\ A=\pm\lf( {k \atop 0}\ {j\atop p}\ri)\ \ ,\ \ 0\leq j<k\ ,\
p\neq 0\ , \cr
I_2&:\ A=\lf( {0 \atop 0}\ {j\atop p}\ri)\ \ ,\ \ (j,p) \neq
(0,0)\ .}}
Clearly, $I_0$ does not give any contribution.
The remaining $\tau$--integrals $I_1$
and $I_2$ are presented in appendix B and we evaluate
for \ftt:
\eqn\fttfinal{
{\cal I}_{2,-2}=8\pi^2 \p_T\lf(\p_T-{2\o T-\ov T}\ri)f+8\pi^2
{(U-\ov U)^2\o (T-\ov T)^2} \p_{\ov U}\lf(\p_{\ov U}+{2\o U-\ov
U}\ri)\ov f\ .}
It is quite remarkable, how e.g. the cubic term of the prepotential
\fone, in the combination of \fttfinal, gives the last term of (B.19).


\subsubsec{4.1.2\ $f_{UU}$}

Similary, an expression with modular weights $(w_T,w_U)=(-2,2)$ can be
found:
\eqn\fuu{
{\cal I}_{-2,2}:=
{(T- \ov T)\o  (U-\ov U)}\ \int {d^2\tau \o \tau_2}\  \sum_{(P_L,P_R)}
P_R^2\   
{\hat Z}_{torus}(\tau,\ov\tau)\ {\ov F}_{-2}(\ov \tau)\ ,} 
which after a Poisson resummation on $m_i$ (cf. Appendix) becomes
\eqn\ftts{
{\cal I}_{-2,2}=
{(T- \ov T)\o  (U-\ov U)}T_2^2
\int {d^2\tau \o \tau_2^4}\ \sum_{n_1,n_2 \atop  l_1,l_2}
Q_R^2\  e^{-2\pi i \ov T \det A}  e^{{-{\pi T_2} \o {\tau_2
U_2}}\lf|n_1\tau+n_2U\tau -Ul_1+l_2 \ri|^2} 
 \ {\ov F}_{-2}(\ov \tau)\ .}
After the integration we end up with:
\eqn\fuufinal{
{\cal I}_{-2,2}=8\pi^2 \p_U\lf(\p_U-{2\o U-\ov U}\ri)f+8\pi^2
{(T-\ov T)^2\o (U-\ov U)^2} \p_{\ov T}\lf(\p_{\ov T}+{2\o T-\ov
T}\ri)\ov f\ .}
Alternatively, with mirror symmetry $T\leftrightarrow U$, which induces
the action $P_L\leftrightarrow P_L, P_R\leftrightarrow \ov P_R$ 
on the Narain momenta one may obtain \fuufinal\ from \ftt.

\subsubsec{4.1.3\ $f_{TU}$}

There are several ways to construct from \setup\ $\tau$--integrals
of weights $w_T=0,w_U=0$ which involve at most two Narain momenta insertions.
Let us take 
\eqn\ftuI{
{\cal I}^a_{0,0}:=
\int {d^2\tau \o \tau_2}\  \sum_{(P_L,P_R)}
{\hat Z}_{torus}(\tau,\ov\tau)\ov F_0(\tau,\ov\tau)\ \ ,} 
with ($\hat{E}_2=E_2-{3\o \pi\tau_2}$)
\eqn\Fzero{
\ov F_0(\tau,\ov\tau)=\lambda_1  
{\hat{\bar E}_2\bar E_4\bar E_6\o \bar\eta^{24}}+
\lambda_2{\bar E_6^2\o \bar \eta^{24}}
+\lambda_3{\bar E_4^3\o \bar \eta^{24}}\ }
and we choose $-264\lambda_1-984\lambda_2+744\lambda_3=0$ to avoid
holomorphic anomalies arising from triangle graphs involving two gauge
fields and the K{\"a}hler-- or sigma model connection 
as external legs with massless states running in the loop. In other
words, we want to discard non--harmonic $\ln(T-\ov T)(U-\ov U)$ terms.
Later we will see that this combination is precisely related to the 
`physical' choice \Fanti\ for
$(\lambda_1,\lambda_2,\lambda_3)=(1/6,1/3,1/2)$. In fact:
\eqn\diff{
{1\o 2\pi i}{\p \o \p\ov\tau}\ov F_{-2}(\ov\tau)=
{1\o 6}{\bar E_2\bar E_4\bar E_6\o \eta^{24}}+
{1\o 3}{\bar E_6^2\o \bar\eta^{24}}
+{1\o 2}{\bar E_4^3\o \bar \eta^{24}}\ .}
The expression ${\cal I}^a_{0,0}$ gives a 
representation for a weight zero automorphic form. However,
non--harmonic, because of $\hat{E}_2$ in \diff. Therefore the theorem
of Borcherds \borch\ does not apply. 
Using results of \hm\ it easily can be
evaluated\foot{In \hm\ the N=2 Green--Schwarz term $G^{(1)}$ is
denoted by $\triangle_{\rm univ}$.}:
\eqn\ftuIres{
{\cal I}^a_{0,0}=16\pi^2 \re(f_{TU})+2G^{(1)}\ .}
Using the explicit form of the N=2 version of the GS--term \hm
\eqn\GS{
G^{(1)}={32\pi^2 \o  (T-\ov T)(U-\ov U)} 
\re\lf\{f-\h(T-\ov T)\p_Tf-\h (U-\ov U) \p_U f\ri\}\ ,}
we arrive at

\eqn\final{
\re\lf\{ D_T D_U f \ri\}={1 \o 16 \pi^2} \int 
{d^2\tau \o \tau_2}\  \sum_{(P_L,P_R)} {\hat Z}_{torus}(\tau,\ov\tau)\ \lf[
{1\o 6}{\hat{\bar E}_2\bar E_4\bar E_6\o \bar\eta^{24}}+{1\o 3}
{\bar E_6^2\o \bar \eta^{24}}+{1\o 2}{\bar E_4^3\o \bar
\eta^{24}}\ri]\ .}

In addition, we want to investigate the integral
\eqn\ftuII{
{\cal I}^b_{0,0}:=
\int {d^2\tau \o \tau_2}\  \sum_{(P_L,P_R)}
\lf(|P_R|^2-{1 \o 2 \pi \tau_2}\ri)\   
{\hat Z}_{torus}(\tau,\ov\tau)\ {\ov F}_{-2}(\ov \tau)\ ,}
The additional GS--like term is needed to
guarantee modular invariance. That can be seen after performing
the Poisson resummation, which yields:

\eqn\ftus{
{\cal I}^b_{0,0}=-T_2^2
\int {d^2\tau \o \tau_2^4}\ \sum_{n_1,n_2 \atop  l_1,l_2}
Q_R \ov Q_R\  e^{-2\pi i \ov T \det A}  e^{{-{\pi T_2} \o {\tau_2
U_2}}\lf|n_1\tau+n_2U\tau -Ul_1+l_2 \ri|^2} 
 \ {\ov F}_{-2}(\ov \tau)\ .}
Again, this integral can be determined using formulas of the appendix
B with the result:  
\eqn\ident{
{\cal I}^a_{0,0}={\cal I}^b_{0,0}\ .}

Let us now come to the identity \diff, which is the link to 
\ident. Rewriting ${\cal I}^b_{0,0}$ as
\eqn\tauder{
{\cal I}^b_{0,0}={i \o \pi}\int {d^2\tau \o \tau_2^2}\ \p_{\ov \tau}(\tau_2 
{Z}_{torus})\ {\ov F}_{-2}(\ov\tau)}
and using \diff\ we may also deduce \ident\ after partial integration.


\subsec{More than two momenta insertions}
Let us give some representative examples.

\subsubsec{4.2.1.\ $f_{TTT}$}

We want to consider the integral
\eqn\fttt{
{\cal I}_{4,-2}:=
{(U- \ov U) \o  (T-\ov T)^2}\ \int d^2\tau
\ \sum_{(P_L,P_R)} P_L \ov P_R^3\ {\hat Z}_{torus}(\tau,\ov\tau)
\ {\ov F}_{-2}(\ov \tau)\ .}
With the identity
\eqn\ida{
P_L\ov P_R^3{\hat Z}_{torus}={{T-\ov T}\o{2\pi \tau_2}}\ \ov P_R^2\ \p_T {\hat 
Z}_{torus}}
we are able to `transform' \fttt\ into a two--momentum integral
of the kind we have discussed before. In particular, this identity tells us
\eqn\covid{
2\pi {\cal I}_{4,-2}=\lf(\p_T+{2\o T-\ov T}\ri){\cal I}_{2,-2}\ .}
Using \fttfinal\ we
obtain after some straightforward algebraic manipulations:
\eqn\ftttfinal{  
{\cal I}_{4,-2}=4\pi f_{TTT}\ .}
This identity was already derived in \afgnt, however in a 
quite different manner.
Moreover, we also may directly integrate \fttt\ as we have done so in the
last subsections.
After a Poisson resummation [cf. (A.1) and 
(A.5)] the integral \fttt\ becomes : 
\eqn\FTTT{\eqalign{
{\cal I}_{4,-2}&=
-{(U- \ov U) \o  (T-\ov T)^2}\ T_2^4\ \int {d^2\tau \o \tau_2^5}\ 
\sum_{n_1,n_2 \atop  l_1,l_2} Q_L\ov Q_R^3\  e^{-2\pi i \ov T \det A}  
e^{{-{\pi T_2} \o {\tau_2 U_2}}\lf|n_1\tau+n_2U\tau -Ul_1+l_2 \ri|^2} 
\ {\ov F}_{-2}(\ov \tau)\cr
&+{3 \o 2\pi}\ {(U- \ov U) \o  (T-\ov T)^2}\ T_2^2 
\int {d^2\tau \o \tau_2^4}\ 
\sum_{n_1,n_2 \atop  l_1,l_2} 
\ov Q_R^2\  e^{-2\pi i \ov T \det A}  e^{{-{\pi T_2} \o {\tau_2 U_2}}
\lf|n_1\tau+n_2U\tau -Ul_1+l_2 \ri|^2} 
 \ {\ov F}_{-2}(\ov \tau)\ .}}
The second integral is of the kind \ftts. 
In fact, using the results of appendix B, we have explicitly
evaluated the integrals 
\FTTT\ and checked \ftttfinal.

\subsubsec{4.2.2.\ $f_{TTU}$}

A covariant expression ${\cal I}_{2,0}$ containg $f_{TTU}$ may be found
by considering $\p_T {\cal I}_{0,0}$:
\eqn\fttu{
{\cal I}_{2,0}:=\p_T {\cal I}^b_{0,0}=
{2\pi\o  (T-\ov T)}\ \int d^2\tau
\  \sum_{(P_L,P_R)} \lf(P_L P_R \ov P_R^2-{P_L\ov P_R\o \pi\tau_2}\ri)   
{\hat Z}_{torus}(\tau,\ov\tau)\ {\ov F}_{-2}(\ov \tau)\ .}
Whereas in \fttu\ each term alone already has weights $(w_T,w_U)=(2,0)$  
and $(w_{\ov T}, w_{\ov U})=(0,0)$, only their combination gives rise 
to a modular invariant integrand. This may be seen after doing the Poisson 
transformation:
\eqn\FTTU{\eqalign{
{\cal I}_{2,0}&=
{2\pi \o  (T-\ov T)}\ T_2^4\ \int {d^2\tau \o \tau_2^5}\ 
\sum_{n_1,n_2 \atop  l_1,l_2} Q_LQ_R\ov Q_R^2\  e^{-2\pi i \ov T \det A}  
e^{{-{\pi T_2} \o {\tau_2 U_2}}\lf|n_1\tau+n_2U\tau -Ul_1+l_2 \ri|^2} 
\ {\ov F}_{-2}(\ov \tau)\cr
&-\ {2\o  (T-\ov T)}\ T_2^2 
\int {d^2\tau \o \tau_2^4}\ 
\sum_{n_1,n_2 \atop  l_1,l_2} 
Q_R\ov Q_R\  e^{-2\pi i \ov T \det A}  e^{{-{\pi T_2} \o {\tau_2 U_2}}
\lf|n_1\tau+n_2U\tau -Ul_1+l_2 \ri|^2} 
\ {\ov F}_{-2}(\ov \tau)\ .}}
Again, for the integration we use the formulas of appendix B
to arrive at:
\eqn\fttufinal{\eqalign{
{\cal I}_{2,0}&=
8\pi^2 \p_T\lf(\p_T-{2\o T-\ov T}\ri)\lf(\p_U-{2\o U-\ov U}\ri)f
-{16\pi^2\o (T-\ov T)^2}\lf(\p_{\ov U}+{2\o U-\ov U}\ri)\ov f\cr
&=8\pi^2 f_{TTU}+2 G_T^{(1)}\ .}}
The second term in the integrand of \fttu\
might be identified with $4 G_T^{(1)}$, although a splitting of both terms
does not make sense because of modular invariance. Besides, only the
combination $8\pi^2 f_{TTU}+2 G_T^{(1)}$ can be written 
covariant w.r.t. \covsT.  

\subsubsec{4.2.3.\ $f_{TUU}$}

Finally, for the integral
\eqn\ftuu{
{\cal I}_{0,2}:={2\pi\o (U- \ov U)}\ \int d^2\tau
\  \sum_{(P_L,P_R)}\lf(P_L \ov P_R P_R^2-{P_L P_R\o \pi\tau_2}\ri) \   
{\hat Z}_{torus}(\tau,\ov\tau)\ {\ov F}_{-2}(\ov \tau)\ ,}
we borrow the results of section (4.2.2) and use mirror symmetry
$T\leftrightarrow U$ to obtain:
\eqn\ftuufinal{\eqalign{
{\cal I}_{0,2}&=
8\pi^2 \p_U\lf(\p_U-{2\o U-\ov U}\ri)\lf(\p_T-{2\o T-\ov T}\ri)f
-{16\pi^2\o (U-\ov U)^2}\lf(\p_{\ov T}+{2\o T-\ov T}\ri)\ov f\cr
&=8\pi^2 f_{TUU}+2 G_U^{(1)}\ .}}

\newsec{Six dimensional origin of the gauge couplings}

Let us consider the amplitudes discussed in the section 3 
from a more general point of view. The gauge kinetic
terms \egaugea\ are deduced from the Einstein term in six
dimensions upon dimensional reduction. In the Einstein
frame the latter does not receive any loop corrections neither in
$d=6$ nor in $d=4$. The relevant object to consider is a two graviton 
amplitude ($i=1,\ldots,6$)
\eqn\six{
\la :\eps_{ij}\bar\p X_1^i\Big[\p X_1^j+i(k_1\cdot\psi_1)\psi_1^j\Big]
e^{ik_1\cdot X_1}: 
:\eps_{kl}\bar\p X_2^k
\Big[\p X_2^l+i(k_2\cdot\psi_2)\psi_2^l\Big] e^{ik_2\cdot X_2}: \ra\ ,}
which may contain both ${\cal R}$ and 
${\cal R}_{ikjl}{\cal R}^{ikjl}$ corrections.
Here $\eps_{ij}$ is the gravitational polarization tensor in six 
dimensions.  
The amplitude is determined by expanding the elliptic genus 
${\cal A}$ of $K3$
w.r.t. ${\cal R}^2$. In general, in N=1,$d=6$ heterotic string theories 
only the $4$--form part of the elliptic genus gives rise to modular
invariant one--loop corrections \lsw\WL.
For the choice of instanton numbers
$(n_1,n_2)=(24,0)$ w.r.t. an $SU(2)$ gauge bundle which leads
to the gauge group $E_7\times E_8'$ this expansion is given by \lsw\WL\hm
\eqn\genus{
\lf. {\cal A}(\tau,{\cal F},{\cal R})\ri|_{4-form}\sim\lf[
({\cal R}^2-{\cal F}^2){\hat{\bar E}_2 \bar E_4\bar E_6 \o
\bar\eta^{24}}+{\bar E_6^2\o \bar\eta^{24}}
{\cal F}^2_{E_7}+{\bar E_4^3\o \bar\eta^{24}}{\cal F}^2_{E'_8}\ri]\ ,}
To saturate the fermionic zero modes one has to contract the four
fermions which is already of the order $\Oc(k^2)$.
Since the worldsheet integral of the bosonic contraction 
$\la \bar \p X^i_1\bar \p X_2^k\ra$ gives a zero result and thus the
$\Oc(k^2)$ term of the effective action vanishes, 
the next to leading order arises from contractions of $\bar \p X^i$ with
the exponential $e^{i k\cdot X}$ which contributes another $\Oc(k^2)$
term to the amplitude. For the $\Oc(k^4)$ contribution we thus obtain \WL:
\eqn\sixDelta{
\triangle_{{\cal R}^2}^{6d,N=1}\sim \int {d^2\tau \o \tau_2^2}
\lf(\bar E_2-{3\o \pi \tau_2}\ri) {\bar E_4 \bar E_6 \o \bar\eta^{24}}
=-8\pi\ .}
The $\hat{\bar E}_2$--piece arises from  
the worldsheet integral over the contractions of $6d$ space--time
bosons:
\eqn\contra{
\int d^2 z_{12}\la\ov\p\bar X_1^i X_2^m\ra\la\ov\p\bar X_2^k X_1^n\ra=
-\eta^{im}\eta^{kn}\int\Big[\bar\p
G_B(z_{12})\Big]^2=\eta^{im}\eta^{kn}{\pi^2\tau_2\o 8}\Big(\bar
E_2-{3\o\pi\tau_2}\Big)\ .}
The appearance of $\hat{\bar E}_2$ in \sixDelta\ may be also understood as
the gravitational charge
$Q_{\rm grav}^2=-2\p_\tau\ln\eta(\tau)=1/6\pi i\  E_2(\tau)$ \agnt.
After the torus compactification we obtain in $d=4$ \WL\agnt\hm
\eqn\fourDelta{
\triangle_{{\cal R}^2}^{4d,N=2}\sim\int {d^2\tau \o \tau_2}{Z}_{torus} 
\lf(\bar E_2-{3\o \pi \tau_2}\ri){\bar E_4 \bar E_6 \o \bar\eta^{24}}\ .}
There is however a subtlety w.r.t. to the indices $i,j,k,l$ 
in deducing the field theoretic kinematic content of \six\ 
for the four dimensional action:
Taking all $i,j,k,l$ as $d=4$ space--time indices $\mu,\nu$ gives
\sixDelta\ for the $\Rc_{\mu\nu\rho\sigma}\Rc^{\mu\nu\rho\sigma}$
correction in $4d$, i.e. \fourDelta\
after taking into account the zero modes of the internal torus.
However when we want to deduce the gauge kinetic term \stgauge,
we keep both $i$ and $k$ as internal indices $+$ which has been
defined in section 3 and 
extract the ${\cal O}(k^2)$--piece of \six\ [cf. \ig]:
\eqn\extract{
\int d^2 z_{12}\la \bar\p X_1^+\bar\p X_2^+\ra=
 2\pi^2 \tau_2 \bar{P}_R^2\ .}
This way we end up at
\eqn\fourF{\Delta_{(TT)}\sim\int {d^2\tau \o \tau_2}
\sum_{(P_L,P_R)}\hat{Z}_{torus} \bar P_R ^2 
{\bar E_4 \bar E_6 \o \bar\eta^{24}}\ .}
after taking into account all kinematic possibilities, in agreement
with \att.
Of course, the $\ov E_2$--part of \sixDelta, measuring the gravitational 
charge, does not occur in $\triangle_{(TT)}$.

\newsec{Conclusion}

We have calculated the one--loop threshold corrections
to the gauge couplings of $U(1)$ gauge bosons which
arise from heterotic N=2,\ $d=4$ compactifications on a torus.
These results fit into the framework
of the underlying N=2 supergravity theory.
Using (4.21) and \nsii\ we may cast the effective gauge couplings
\atu\ into the final form:

\eqn\finalgauge{
-{\triangle_{(TU)}\o 32\pi^2}={1\o 2} \re\lf\{ D_T D_U f\ri\}
={1\o 16\pi^2} \lf[\ln|j(T)-j(U)|^2-G^{(1)}+\sigma(T,U)\ri]\ ,}
where $\sigma(T,U)$ are the universal one--loop corrections
appearing in all gauge threshold corrections of heterotic
N=2 theories \nsii. The correction $G^{(1)}$ describes the
mixing of the dilaton and the moduli fields at one--loop \jv\dewit\nsii.

In section 4 we have calculated several world--sheet $\tau$--integrals
as they appear in string amplitude calculations from various
contractions of the internal bosonic coordinate fields.
These expressions appear quite general in 
heterotic torus compactifications from N=1 in $d=10,6,4$ to $d=8,4,2$.
The relevant string amplitudes take a generic form, given by a 
$\tau$--integral over the (new)
supersymmetric index \susyindex\ (or variants of it depending on $d$), 
completed with momentum insertions of internal
bosonic fields, which take into account either vertex operator contractions
or charge insertions. In the case of $K3\times T^2$ compactifications,
a part of the supersymmetric index \susyindex\ is related to a modular
function $f$, which is the N=2, $d=4$ prepotential. Many
$\tau$--integrals can be expressed through $f$ and its derivatives.
Such relations between string--amplitudes and a function $f$ and its
derivatives, as established here for N=2 in $d=4$, should also exist 
in any torus compactifications of e.g. $d=10,4$ heterotic string theory.

\smallskip

\ \ 
{\bf Acknowledgments:}
We are very grateful to I. Antoniadis, J.--P. Derendinger and W. Lerche 
for important discussions.  
K.F. thanks the Institut de Physique Th{\'e}orique de 
l'Universit{\'e} de Neuch{\^a}tel for the friendly hospitality.
This work is supported by the
Swiss National Science Foundation, and the European Commission TMR programme 
ERBFMRX--CT96--0045, in which K.F. is associated to HU Berlin and St. St.
to OFES no. 95.0856.

\appendix{A}{\bf Poisson resummation}

\def\bp {{\vec p }}  \def\by {{\vec y}} \def\bb {{\vec b}} \def\bc {{\vec c}} 

In this section we want to perform a Poisson transformation on
the expression
\eqn\start{
{\cal S}=\sum_{\bp\in \Lambda^\ast} e^{-\pi \bp^t \alpha \bp} 
e^{2\pi i \by^t \bp}(\bp^t A \bp+\bb^t \bp+a_0)(\bp^t D \bp+\bc^t \bp+e_0)\ ,}
for some matrices $\alpha,A,D$ ($\det\alpha\neq 0$), 
vectors $\by,\bb,\bc$ 
and scalars $a_0,e_0$.
This is achieved like one does a Fourier transformation on a periodic
function $F(\vec x)$, i.e. $[F(\vec x+\vec p_0)=F(\vec x)]$:
\eqn\fourier{
F(\vec x)=\sum_{\bp \in \Lambda} e^{-\pi (\bp+\vec x)^t\alpha(\bp+\vec
x)} e^{2\pi i \by^t(\bp+\vec x)}\ .}
With 
\eqn\dual{
F^{\ast}(\vec q)={1\o vol(\Lambda)}\int\limits_{-\infty}^\infty 
d \vec x\  e^{-2\pi i \vec q^t \vec x} F(\vec x)\ ,}
we may write:
\eqn\ffinal{\eqalign{
F(\vec x)&=\sum_{\vec q\in \Lambda^\ast}\  e^{2\pi i {\vec q}^t \vec x}
F^\ast(\vec q)\cr
&={1\o \sqrt{\det \alpha}}{1\o vol(\Lambda)}
\sum_{\vec q \in \Lambda^\ast} e^{-2\pi i {\vec q}^t \vec x}
e^{-\pi (\vec y+\vec q)^t\alpha^{-1}(\vec y+\vec q)}\ .}}
Along that way a Fourier transformation on \start\ yields

\eqn\resfour{\eqalign{
{\cal S}&={1\o \sqrt{\det \alpha}}{1\o vol(\Lambda^\ast)}
          \sum_{\vec q \in \Lambda} 
          e^{-\pi (\vec y+\vec q)^t\alpha^{-1}(\vec y+\vec q)}\ 
          \lf\{a'_0e'_0+{1\o 2\pi}\Big[e'_0 \tr(\alpha^{-1}A)
          +a'_0 \tr(\alpha^{-1}D)\ri.\cr
        &+\lf(\vec b+i(A^t+A)\alpha^{-1}(\vec y+\vec q)\ri)^t  \alpha^{-1}
          \Big(\vec c+i(D^t+D)\alpha^{-1}(\vec y+\vec q)\Big)  \Big]\cr
        &\lf. +{1\o 4\pi^2}\Big[\tr(\alpha^{-1}A)\tr(\alpha^{-1}D)+
          \tr(\alpha^{-1}A\alpha^{-1}D^t)+
           \tr(\alpha^{-1}A\alpha^{-1}D) \Big] \ri\}  }}
with the following abbreviations:
\eqn\abbrev{\eqalign{
a_0'&=a_0-(\vec y+\vec q)^t\alpha^\ast A\alpha^{-1}(\vec y+\vec q)+
i\vec b^t\alpha^{-1}(\vec y+\vec q)\cr
e_0'&=e_0-(\vec y+\vec q)^t\alpha^\ast D\alpha^{-1}(\vec y+\vec q)+
i\vec c^t\alpha^{-1}(\vec y+\vec q)\ .}}

\appendix{B}{\bf Integrals}

\subsec{Orbit $I_1$}

For the orbit $I_1$ we have to face the following integrals
\eqn\genericint{
I_1^{\alpha,\beta,n}:=\sum_{k,j,p}{\tilde I}_1^{\alpha,\beta,n}=
T_2\ e^{-2\pi i \ov Tkp}
\int\limits_{H_+} {d\tau \o \tau_2^{2+\beta}} \sum_{k,j,p}
\tau_1^\alpha e^{-{\pi T_2\o \tau_2 U_2}|k\tau-j-pU|^2}
e^{-2\pi i\ov\tau n}\ ,}
for the cases $\alpha=0,\ldots,4$ and $\beta=-1,0,\ldots,3$.
Clearly, the case $\alpha=0$ and $\beta=0$ corresponds to
the integral performed in \DKLII. The case $\beta=-1$ is needed for
the integrals appearing in sect. 4.2.
We expanded the anti--holomorphic function $\ov F(\ov \tau)$ in a power
series:
\eqn\series{
\ov F(\ov \tau)=\sum_{n \geq -1} c_n e^{-2\pi i\ov\tau n}\ .}
The $\tau_1$--integration of \genericint\ can be reduced to Gaussian
integrals:
\eqn\zwischen{\eqalign{
{\tilde I}_1^{\alpha,\beta,n}&={k^{-\alpha}\o k}\sqrt{T_2U_2}\ e^{-2\pi i \ov
Tkp}\ 
e^{2\pi T_2 kp}\ e^{-2\pi i {n \o k}(j+pU_1)}\cr
&\times\int_0^\infty {d\tau_2 \o 
\tau_2^{{3\o 2}+\beta}}\ e^{-{\pi T_2 \o U_2}(k+{nU_2 \o k T_2})^2\tau_2}
\ e^{-\pi p^2T_2 U_2/\tau_2}\ {\cal X}_\al\ ,}}
with: 
\eqn\intalpha{\eqalign{
{\cal X}_0&=1\cr
{\cal X}_1&=-i{n\o k} {\tau_2 U_2 \o T_2}+j+pU_1 \cr
{\cal X}_2&={1\o 2\pi}{\tau_2 U_2 \o T_2} + 
(-i{n\o k} {\tau_2U_2 \o T_2}+j+pU_1)^2\cr
{\cal X}_3&={3\o 2\pi} {\tau_2 U_2\o T_2}
(-i{n\o k} {\tau_2U_2 \o T_2}+j+pU_1)+
(-i{n\o k} {\tau_2U_2 \o T_2}+j+pU_1)^3\cr
{\cal X}_4&={3\o 4\pi^2} \lf({\tau_2 U_2\o T_2}\ri)^2+
{3\o\pi}{\tau_2 U_2\o T_2}(-i{n\o k} {\tau_2U_2 \o T_2}+j+pU_1)^2+
(-i{n\o k} {\tau_2U_2 \o T_2}+j+pU_1)^4\ .}}
Next, we have to do the $\tau_2$--integration. For $\beta=0,1,2$ 
we may borrow results\foot{We thank N.A. Obers for explanation of the notation
of \bachas. In their final formulae
one must replace $p\rightarrow |p|$.}
from \bachas\ for the integrals
\eqn\zwischenii{
{\tilde I}_1^{0,\beta,n}=\lf(k\o |p| U_2\ri)^\beta {\tilde I}_1 
                     \times \cases{\lf[1+n{{\cal U}_2\o{\cal
                     T}_2}\ri]^{-1} &$\beta=-1$\ ,\cr
                     1 &$\beta=0$\ ,\cr
                     \lf[1+{1\o {\cal T}_2}(n{\cal U}_2+{1\o 2\pi})\ri]
                     &$\beta=1$\ ,\cr
                     \lf[1+{1\o {\cal T}_2}(2n{\cal U}_2+{3 \o 2\pi})
                     +{1\o {\cal T}_2^2}(n^2{\cal U}_2^2+{3n{\cal U}_2
                     \o 2\pi}+{3 \o 4 \pi^2})\ri]&$\beta=2$\ , \cr
                     \lf[1+{1\o {\cal T}_2}(3n{\cal U}_2+{3 \o \pi})
                     +{1\o {\cal T}_2^2}(3n^2{\cal U}_2^2+{6n{\cal U}_2
                     \o \pi}+{15 \o 4 \pi^2})\ri.&$\ $\cr
                     \lf.\ \ \ \ +{1\o {\cal T}_2^3}(n^3{\cal U}_2^3+
                     {3n^2\o\pi}{\cal U}_2^2+{15\o 4 \pi^2}n {\cal U}_2+
                     {15\o 8\pi^3})\ri]&$\beta=3$\ , \cr}}
with
\eqn\zwischeniii{
\tilde I_1={1\o k|p|}e^{-2\pi i\ov T kp} e^{-2\pi i {n \o k}(j+pU_1-i|p|U_2)}
\ e^{2\pi T_2 k(p-|p|)}\ ,}
and: 
\eqn\moduli{
{\cal T}_2=k|p|T_2\ \ \ ,\ \ \ {\cal U}_2={|p|U_2 \o k}\ .}
Before we continue, let us recover 
\eqn\easy{\eqalign{
I_1^{0,0,n}&=\sum_{k>0 \atop l\in \IZ}
\delta_{n,kl}\ \Li_1\lf[e^{2\pi i(kT+lU)}\ri]+hc.\ , \cr
I_1^{0,1,n}&=\sum_{k>0 \atop l\in \IZ}
\delta_{n,kl}\ {\cal P}\lf[e^{2\pi i(kT+lU)}\ri]+hc.\ ,}}
with:
\eqn\calp{
{\cal P}(kT+lU)=\im(kT+lU)\ \Li_2\lf[e^{2\pi i(kT+lU)}\ri]+{1\o 2\pi}
\Li_3\lf[e^{2\pi i(kT+lU)}\ri]\ .}

Finally for the cases of interest, 
we have reduced everything to integrals
${\tilde I}_1^{0,\beta,n}$ given in eq. \zwischenii .
\eqn\last{
\kern-1em 
\eqalign{
k \tilde I_1^{1,\beta,n}&=-i{n\o k} {U_2 \o T_2}{\tilde I}_1^{0,\beta-1,n}
+(j+pU_1){\tilde I}_1^{0,\beta,n}\cr
k^2 {\tilde I}_1^{2,\beta,n}&=-{n^2\o k^2} {U_2^2 \o T_2^2}
{\tilde I}_1^{0,\beta-2,n}+{1\o 2\pi}{U_2 \o T_2}{\tilde I}_1^{0,\beta-1,n}
-2i(j+pU_1){n\o k} {U_2 \o T_2}{\tilde I}_1^{0,\beta-1,n}\cr
&\ \ \ +(j+pU_1)^2{\tilde I}_1^{0,\beta,n}\cr
k^3 {\tilde I}_1^{3,\beta,n}&=i{n^3\o k^3}{U_2^3\o T_2^3}
{\tilde I}_1^{0,\beta-3,n}-{3\o 2\pi}i{n\o k}{U_2^2\o T_2^2}
{\tilde I}_1^{0,\beta-2,n}-3{n^2\o k^2}(j+pU_1){U_2^2\o T_2^2}
{\tilde I}_1^{0,\beta-2,n}\cr
&\ \ \ +{3\o 2\pi}(j+pU_1)
{U_2\o T_2}{\tilde I}_1^{0,\beta-1,n}
-3i(j+pU_1)^2{n\o k}{U_2\o T_2}{\tilde I}_1^{0,\beta-1,n}+(j+pU_1)^3
{\tilde I}_1^{0,\beta,n}\cr
k^4 {\tilde I}_1^{4,\beta,n}&=
{n^4\o k^4}{U_2^4\o T_2^4}{\tilde I}_1^{0,\beta-4,n}
-{3\o \pi}{n^2\o k^2}{U_2^3\o T_2^3}{\tilde I}_1^{0,\beta-3,n}
+4i{n^3\o k^3}{U_2^3\o T_2^3}(j+pU_1){\tilde I}_1^{0,\beta-3,n}\cr
&\ \ \ -{6\o \pi}i{n\o k}(j+pU_1){U_2^2\o T_2^2}{\tilde
I}_1^{0,\beta-2,n}+{3\o 4\pi^2}{U_2^2\o T_2^2}{\tilde I}_1^{0,\beta-2,n}
-6{n^2\o k^2}{U_2^2 \o T_2^2}(j+pU_1)^2{\tilde I}_1^{0,\beta-2,n}\cr
&\ \ \ +{3\o \pi}{U_2\o T_2}(j+pU_1)^2{\tilde I}_1^{0,\beta-1,n}
-4i{n\o k}{U_2\o T_2}(j+pU_1)^3{\tilde I}_1^{0,\beta-1,n}
+(j+pU_1)^4{\tilde I}_1^{0,\beta,n}\ .}}
Now, the important and nice fact is, that after expanding the
expression
\last, the $j$--sum, in the combination of (4.10), (4.16), (4.25),
(4.32) and (4.34) becomes trivial and gives the restriction 
$n=kl\ ,\ l\in \IZ$.  
Then, after a straightforward calculation the orbits 
$I^{{\cal I}_{w_T,w_U}}_1$ belonging to the integral ${\cal
I}_{w_T,w_U}$ can be determined:
\eqn\IIZ{\eqalign{
I^{{\cal I}_{2,-2}}_1&=-\sum_{k>0\atop l\in \IZ}\sum_{p>0}
\delta_{n,kl}\lf(2{k^2\o p}+{k\o \pi T_2  p^2 }+{1 \o 4\pi^2 T_2^2 p^3 }\ri)
x^p\cr
&\ \ \ -\sum_{k>0\atop l\in \IZ}
\sum_{p>0}\delta_{n,kl}\ \lf(2 {U_2^2 l^2 \o T_2^2 p}+{U_2 l\o \pi T_2^2 p^2}+
{1\o 4\pi^2 T_2^2 p^3}\ri)\ov x^p\cr
I^{{\cal I}_{0,0}}_1&=-\sum_{k>0\atop l\in \IZ}\sum_{p>0}
\ \delta_{n,kl}\lf({2kl\o p}+{l\o \pi T_2 p^2}+
{k\o \pi U_2 p^2}+{1\o 2\pi^2 T_2 U_2 p^3}\ri)x^p+hc.\cr
I^{{\cal I}_{4,-2}}_1&=-2i\sum_{k>0\atop l\in \IZ}\sum_{p>0}\delta_{n,kl}\ 
k^3 x^p\cr
I^{{\cal I}_{2,0}}_1&=-\sum_{k>0\atop l\in \IZ}\sum_{p>0}\ \delta_{n,kl}
\lf(4i\pi k^2l+{2ik^2\o U_2 p}+{ik\o \pi T_2 U_2p^2}+{2ikl\o T_2 p}
+{i\o \pi^2 T_2^2 U_2p^3}+{il\o\pi T_2^2p^2}\ri)x^p\cr
&\ \ \ -\sum_{k>0\atop l\in \IZ}\sum_{p>0}\ \delta_{n,kl}
\lf({i\o 4\pi^2T_2^2 U_2p^3}-{il\o 2\pi T_2^2 p^2}\ri)\ov x^p\ ,}}
with $x:=e^{2\pi i(kT+lU)},\ \ov x:=e^{-2\pi i(k\ov
T+l\ov U)}$.

\subsec{Orbit $I_2$}

For the orbit $I_2$ the following integrals appear
\eqn\Ithree{
I^{\alpha,\beta,\gamma,n}_2=T_2\int_{-\h}^{+\h}d\tau_1{\int_0^\infty}
{d\tau_2 \o \tau_2^{2+\gamma}}\sum_{(j,p)}'j^\alpha p^\beta\ 
e^{-{\pi T_2 \o \tau_2 U_2}|j+Up|^2} e^{-2\pi i\ov\tau n}\ .}
Here the prime at the sum means that we do not sum over 
$(j,p)=(0,0)$, which is taken into account in $I_0$. The case
$\gamma=0$ describes the respective integral of \DKLII. In that case one 
has to regularize the integral. We will only need cases with
$\gamma\neq 0$. Therefore we have not to introduce an IR--regulator.
See also \kk\ for discussions. 
This also means that our results will not produce any non--harmonic
$\ln(T_2 U_2)$--pieces. Such terms are usually signals of potential 
anomalies arising in the IR.
The $\tau_1$--integration is trivial and simply projects onto massless
states, i.e. $n=0$. For the $\tau_2$--integration we again 
may use results of \bachas:
\eqn\intbach{
\int_0^\infty{d\tau_2\o \tau_2^{2+\gamma}} \sum_{(j,p)}'\ 
e^{-{\pi T_2 \o \tau_2 U_2}|j+Up|^2}=\Gamma(\gamma+1)\lf({U_2\o \pi
T_2}\ri)^{\gamma+1}\ \sum_{(j,p)}^{'}{1\o |j+p U|^{2\gamma+2}}\ .}
Therefore, we only have to concentrate on the sum
\eqn\sommer{
\sum_{(j,p)}'{j^\alpha p^\beta\o |j+Up|^{2\gamma+2}}\ .}

Let us first perform the summation for the case $p\neq 0$ and write:
\eqn\Zwisch{\eqalign{
\sum_{p\neq 0}{j^\alpha p^\beta\o
[(j+U_1p)^2+p^2U_2^2]^{\gamma+1}}&=
{1\o\gamma !}\sum_{p>0}\lf({1\o i} {\p\o \p\theta}\ri)^\alpha p^{\beta-2\gamma}
\lf({(-1)\o 2U_2} {\p \o \p U_2}\ri)^\gamma\cr
&\times\sum_{j=-\infty}^\infty\lf.\lf[{e^{i\theta j}\o (j+U_1p)^2+p^2U_2^2}+
{(-1)^\beta\ e^{i\theta j}\o (j-U_1p)^2+p^2U_2^2}\ri]
\ri|_{\theta=0}\ .}}
After using a Sommerfeld--Watson transformation, introduced in \hm, 
($C>0,\ 0\leq\theta\leq 2\pi$)
\eqn\watson{
\sum_{j=-\infty}^\infty{e^{i\theta j}\o (j+B)^2+C^2}={\pi \o C}e^{-i\theta
(B-iC)}{1\o 1-e^{-2\pi i(B-iC)}}+{\pi \o C}e^{-i\theta 
(B+iC)}{e^{2\pi i(B+iC)}\o 1-e^{-2\pi i(B+iC)}}\ ,}
we end up with
\eqn\ZW{\eqalign{
&\ {\pi\o \gamma!}\lf({-1\o 2 U_2} {\p \o \p U_2}\ri)^\gamma\Big\{
{1\o U_2}\Big[[(-U)^\alpha+(-1)^\beta U^\alpha]\sum_{l>0}
\Li_{1-\alpha-\beta+2\gamma}(q_U^l)\cr
&\ \ \ +[(-\ov U)^\alpha+(-1)^\beta \ov U^\alpha]
\sum_{l>0}\Li_{1-\alpha-\beta+2\gamma}(\ov q_U^l)\cr
&\ \ \ +[(-\ov U)^\alpha+(-1)^\beta U^\alpha]
\zeta(1-\alpha-\beta+2\gamma)\Big]\Big\}\ .}}

Let us now come to the case $p=0$ of \sommer
\eqn\asdf{
Q^{\alpha,\beta,\gamma}:=\delta_{\beta,0}\ [1+(-1)^\alpha]
\ \sum_{j=1}^\infty {1\o j^{2\gamma+2-\alpha}}\ ,}
which only gives a non--zero contribution for $\beta=0$. 
Moreover, we must only consider $\alpha \in 2\IZ$.
For the examples we discuss in section 4.1. we have $\alpha+\beta=\gamma=2$ 
and for the cases in section 4.2, $\alpha+\beta=4,\ \gamma=3$.
In both cases, the sum \asdf\ becomes:
\eqn\cont{
Q^{2,0,2}=Q^{4,0,3}=2\sum_{j=1}^\infty {1\o j^4}={\pi^4\o 45}\ .}

Putting everything together, we obtain for $I_2^{\alpha,\beta,\gamma,n}$:
\eqn\FINAL{\eqalign{
I_2^{\alpha,\beta,\gamma,n}&=\delta_{n,0}T_2\ \Gamma(\gamma+1)
\lf({U_2\o \pi T_2}\ri)^{\gamma+1}\cr
&\times\Big\{\ {\pi\o \gamma!}\lf({-1\o 2 U_2} {\p \o \p U_2}\ri)^\gamma\Big[
{1\o U_2}\Big([(-U)^\alpha+(-1)^\beta U^\alpha]\sum_{l>0}
\Li_{1-\alpha-\beta+2\gamma}(q_U^l)\cr
&\ \ \ +[(-\ov U)^\alpha+(-1)^\beta \ov U^\alpha]
\sum_{l>0}\Li_{1-\alpha-\beta+2\gamma}(\ov q_U^l)\cr
&\ \ \ +[(-\ov U)^\alpha+(-1)^\beta U^\alpha]
\zeta(1-\alpha-\beta+2\gamma)\Big)\Big]\cr  
&+Q^{\alpha,\beta,\gamma}\Big\}\ .}}

\listrefs
\end


From foerger@pth.polytechnique.fr Thu Nov 13 11:35 MET 1997
Date: Thu, 13 Nov 1997 11:37:53 +0100 (MET)
From: Kristin Foerger <foerger@pth.polytechnique.fr>
To: Stieberger Stephan <Stephan.Stieberger@cern.ch>
Subject: Re: your mail
In-Reply-To: <199711122315.AAA14743@surya11.cern.ch>
MIME-Version: 1.0
Content-Type: TEXT/PLAIN; charset=US-ASCII
Content-Length: 65171

\input harvmac

\def\h {{1\over 2}}
\def\al {\alpha}

\def\ov {\overline}
\def\o {\over}
\def\Li {{\cal L}i}

\def\cmp {{\it Comm. Math. Phys.\ }}
\def\np {{\it Nucl. Phys.\ } {\bf B\ }}
\def\pl {{\it Phys. Lett.\ } {\bf B\ }}
\def\pr {{\it Phys. Rept.\ } }

\def\br{\hfill\break}
\def\IZ{ {\bf Z}}
\def\IF{{\bf F}}
\def\subsubsec #1{\ \br \noindent {\it #1} \br}
\def\Ac {{\cal A}}

\def\Ic {{\cal I}}
\def\Kc {{\cal K}}
\def\Lc {{\cal L}}
\def\Nc {{\cal N}}
\def\Oc {{\cal O}}
\def\Rc {{\cal R}}

\def\th {\theta}

\def\tr {{\rm Tr}}
\def\det {{\rm det}}

\def\lf {\left}
\def\ri {\right}
\def\ra {\rightarrow}

\def\al {\alpha}
\def\re {{\rm Re}}
\def\im {{\rm Im}}
\def\p {\partial}
\def\la{\langle}
\def\ra{\rangle}

\def\eps{\epsilon}

\lref\klt{V. Kaplunoksky, J. Louis and S. Theisen, \pl {\bf 357}
(1995) 71}
\lref\kv{S. Kachru and C. Vafa, \np {\bf 450} (1995) 69}

\lref\higgs {M. Dine, P. Huet and N. Seiberg, \np {\bf 322} (1989) 301;
L. Ibanez, W. Lerche, D. L{\"u}st and S. Theisen, \np {\bf 352} (1991) 435}

\lref\msV {P. Mayr and S. Stieberger, \pl {\bf 355} (1995) 107}
\lref\jv {{\it see e.g.:} J.--P. Derendinger, S. Ferrara, C. Kounnas
and F. Zwirner, \np {\bf 372} (1992) 145;
V. Kaplunovsky and J. Louis, \np {\bf 444} (1995) 191}

\lref\bachas {C. Bachas, C. Fabre, E. Kiritsis, N.A. Obers and
P. Vanhove, hep--th/9707126}
\lref\st {S. Stieberger, NEIP--001/97, work in progress}
\lref\borch {R.E. Borcherds, {\it Invent. Math.} {\bf 120} (1995) 161} 
\lref\WL {W. Lerche, \np {\bf 308} (1988) 102}
\lref\lsw {W. Lerche, B.E.W. Nilsson and A.N. Schellekens, \np {\bf 289}
(1987) 609; W. Lerche, B.E.W. Nilsson, A.N. Schellekens and N.P. Warner,
\np {\bf 299} (1988) 91; W. Lerche, A.N. Schellekens and N.P. Warner, 
\pr {\bf 177} (1989) 1}
\lref\agnt{I. Antoniadis, E. Gava and K.S. Narain, \np {\bf 383}
(1992) 92; \pl {\bf 283} (1992) 209}

\lref\nsii {H.P. Nilles and S. Stieberger, \np {\bf 499} (1997) 3}
\lref\hm {J.A. Harvey and G. Moore, \np {\bf 463} (1996) 315}
\lref\klm {A. Klemm, W. Lerche and P. Mayr, \pl {\bf 357} (1995) 313}

\lref \DKLI {L. Dixon, V. Kaplunovsky and J. Louis, \np {\bf 329} (1990) 27}
\lref \strominger{A. Strominger, \cmp {\bf 133} (1990) 163}
\lref\cdfp {A. Ceresole, R. d' Auria, S. Ferrara and A. van Proeyen,
\np {\bf 444} (1995) 92}

\lref\all {B. de Wit, P.G. Lauwers, 
R. Philippe, S.Q. Su, A. van Proeyen, \pl {\bf 134} (1984) 37; 
B. de Wit, A. van Proeyen, \np {\bf 245} (1984) 89;
B. de Wit, P.G. Lauwers, A. van Proeyen, \np {\bf 255} (1985) 569; 
E. Cremmer, C. Kounnas, A. van Proeyen, J.P. Derendinger, 
S. Ferrara, B. de Wit, L. Girardello, \np {\bf 250} (1985) 385}

\lref \agntz {I. Antoniadis, E. Gava, K. Narain and T. Taylor,
\np {\bf 432} (1994) 187}
\lref \vk {V. Kaplunovsky, \np {\bf 307} (1988) 145 and hep--th/920570}
\lref \DKLII {L. Dixon, V. Kaplunovsky and J. Louis, \np {\bf 355} (1991) 649}

\lref\dewit {B. de Wit, V. Kaplunovsky, J. Louis and D. L{\"u}st, 
\np {\bf 451} (1995) 53}

\lref\afgnt {I. Antoniadis, S. Ferrara, E. Gava, K. Narain and T. Taylor,
\np {\bf 447} (1995) 35}

\lref\kk{E. Kiritsis and  C. Kounnas, 
\np {\bf 442} (1995) 472; P.M. Petropoulos and J. Rizos, \pl 
{\bf 374} (1996) 49}

\lref\dfkz {J.P. Derendinger, S. Ferrara, C. Kounnas and F. Zwirner, 
\np {\bf 372} (1992) 145;
I. Antoniadis, E. Gava, K.S. Narain and T.R. Taylor,
\np {\bf 407} (1993) 706}

\lref\msi {P. Mayr and S. Stieberger, \np {\bf 407} (1993) 725;
D. Bailin, A. Love, W. Sabra and S. Thomas, 
{Mod. Phys. Lett.} {\bf A9} (1994) 67; {\bf A10} (1995) 337}
\lref\min {J. A. Minahan, \np {\bf 298} (1988) 36; P. Mayr,
S. Stieberger, \np {\bf 412} (1994) 502; K. F{\"o}rger, B.A. Ovrut,
S. Theisen, D. Waldram, \pl {\bf 388} (1996) 512} 

\lref\ejm {J. Ellis, P. Jetzer, L. Mizrachi, \np {\bf 303} (1988) 1}
\lref\gkkopp {A. Gregori, E. Kiritsis, C. Kounnas, N. A. Obers, P. M. 
Petropoulos, B. Pioline, hep--th/9708062}
\lref\vadim{ V. Kaplunovsky, \np {\bf 307} (1988) 145}
\lref\llt{J. Lauer, D. L{\"u}st, S. Theisen, \np {\bf 309} (1988) 771}

\Title{\vbox{\rightline{\tt hep-th/9709004} 
\rightline{NEIP--009/97}    }}
{\vbox{\centerline{String Amplitudes and N=2, $d=4$ Prepotential}
\vskip4pt\centerline{in Heterotic $K3\times T^2$ Compactifications }}}

\centerline{K. F{\"o}rger$^1$\ \ and\ \ S. Stieberger$^2$}

\bigskip\centerline{\it $^1$Sektion Physik}
\centerline{\it Universit{\"a}t M{\"u}nchen}
\centerline{\it Theresienstra\ss e 37}
\centerline{\it D--80333 M{\"u}nchen, FRG}

\vskip .2in
\bigskip\centerline{\it $^2$Institut de Physique Th{\'e}orique}
\centerline{\it Universit{\'e} de Neuch{\^a}tel}
\centerline{\it CH--2000 Neuch{\^a}tel, SWITZERLAND}

\vskip .3in
For the gauge couplings, which arise after toroidal compactification
of six--dimensional heterotic N=1 string theories from the $T^2$ torus,
we calculate their one--loop corrections. This is performed by considering 
string amplitudes involving two gauge fields and moduli fields.
We compare our results with the equations following from N=2 special
geometry and the underlying prepotential of the theory.
Moreover we find relations between derivatives of
the N=2, $d=4$ prepotential and world--sheet $\tau$--integrals which appear
in various string amplitudes of any $T^2$--compactification.
\Date{09/97} 


\newsec{Introduction}
The vector multiplet sector of N=2 supergravity in four dimensions is 
governed by a holomorphic function, namely the
prepotential $F$ \all. 
Up to second order derivatives the Lagrangian is constructed from it.
The K{\"a}hler metric and the
gauge couplings are expressible by derivatives of this prepotential.
Effective N=2 supergravity theories in four dimensions arise e.g. after
compactifications of heterotic string theories on $K3\times T^2$
or type IIA or type IIB on a Calabi--Yau (CY) threefold.
In the following we will concentrate on the first vacuum.
In addition to the heterotic gauge group, which depends on the
specific instanton embeddings and may be --in certain cases-- 
completely Higgsed away,
a $K3\times T^2$ heterotic string compactification always possesses
the $U(1)^2_+\times U(1)^2_-$ internal gauge symmetry. The first factor
comes from the internal graviton and the last factor arises from
the compactification of the six--dimensional tensor multiplet which
describes the heterotic dilaton in six dimensions.
The gauge fields of the left $U(1)_L$'s belong to vector multiplets
whose scalars are the $T$ and $U$ moduli of the torus $T^2$.
Their moduli space is described by special geometry \strominger\DKLI. 
Besides, there is a vector multiplet for the heterotic dilaton and
the graviphoton, whereas the $K3$
moduli and the gauge bundle deformations come in scalars of hyper multiplets.

At special points in the $T,U$ moduli space,
the $U(1)_L^2$ may be enhanced to $SU(2)_L\times U(1)_L$ for $T=U$,
to $SU(2)_L^2$ for $T=U=i$ and $SU(3)_L$ at $T=U=\rho$ where
$\rho=e^{2\pi i/3}$ \higgs.
Logarithmic singularities in the effective string
action appear at these special points. 
This effect should be seen e.g. in the one--loop gauge 
couplings, where heavy strings have been integrated out. Modes, which 
have been integrated out and become
light at these special points in the moduli space are responsible
for these singularities.
These one--loop couplings can be expressed by the perturbative
prepotential.
Therefore, a calculation of one--loop gauge couplings gives
information about the prepotential (and vice versa). Until 
now\foot{Except \msV.}, 
no threshold calculation has been undertaken for gauge groups, where
the modulus and the gauge boson under consideration 
sit in the same vector multiplet.
In the following we focus on the perturbative prepotential
of the vector multiplets.

One--loop gauge threshold corrections are very important quantities for 
at least three aspects of string theory: They play an important
r{\^o}le in constructing consistent, i.e. anomaly free, effective string
actions \jv, they are sensitive quantities for heterotic--typeII string
duality tests \kv\klm\klt\ 
and shed light on the perturbative and hopefully also on the
non--perturbative part of the prepotential \hm.
In this paper we will focus on the last issue: We calculate
threshold corrections to the $U(1)_L^2$ gauge bosons which are
given as certain $\tau$--integrals. The latter we can 
express in terms of the prepotential and its derivatives.
This not only gives a
check of the framework of N=2 supergravity emerging for
heterotic $K3\times T^2$ string compactifications, which tells us, how these
corrections have to be expressed by the prepotential, but it gives
relations between string amplitudes given as world--sheet torus
integrals and the prepotential. Such relations are important for
understanding the structure of string amplitudes in any dimensions
and its relation to a function, which in $d=4$ is the N=2 prepotential.

The organization of the present paper is as follows:
In section 2 we briefly review some facts about N=2 supergravity
in light of the gauge couplings. Section 3 is devoted to a
string derivation of the $U(1)_L^2$--gauge couplings by calculating
the relevant string amplitudes.
In section 4 we find various relations between world--sheet 
$\tau$--integrals and combinations of the prepotential and its 
derivatives. These relations allow us to express the string amplitudes
in terms of the prepotential which is the appropriate form
to compare with the supergravity formulae of section 2.
In section 5 we trace back the origin of the gauge couplings 
to six dimensions via the elliptic genus.
Section 6 gives a summary of our results and some concluding
remarks. All technical details for the $\tau$--integrations are
presented in the appendix.

\newsec{N=2 supergravity and effective string theory}

For N=2,\ $d=4$ heterotic string vacua arising from $K3\times T^2$
compactifications, the dilaton field $S$, the $T$ and $U$ moduli
describing the torus $T^2$ (and possible Wilson lines) are scalar fields
of N=2 vector multiplets \cdfp\dewit. The absence of couplings between 
scalars of vector multiplets and scalars of hyper multiplets 
(describing e.g. the K3 moduli) allows one to study the two moduli
spaces seperately \DKLI.
Since the dilaton field $S$ comes in a vector multiplet, the 
prepotential describing the gauge sector may receive  
space--time perturbative and non--perturbative corrections in
contrast to the hyper multiplet moduli space, which does not get any
perturbative corrections.
However in N=2 supergravity one expects a non--renormalization 
theorem following from chiral superspace integrals which prohibits higher than
one--loop corrections to the prepotential.
Besides, in heterotic string theories, the dilaton obeys a
continuous Peccei--Quinn symmetry to all orders in perturbation 
theory which also forbids higher than one--loop corrections.
Therefore, the only perturbative correction to the tree--level prepotential
comes at one--loop, summarized by a function $f$.  
For the prepotential describing the $S,T$ and $U$
moduli space of a heterotic $K3\times T^2$ compactification one has 
 (neglecting non--perturbative corrections) \dewit\afgnt
\eqn\prep{
F(X)={X^1X^2X^3 \o X^0} +(X^0)^2 f\lf({X^2\o X^0},{X^3\o X^0}\ri)\ ,}
with the unconstrained vector multiplets $X^I,\ I=0,\ldots,3$. 
The function $f$ is much constrained by the perturbative duality 
group $SL(2,\IZ)_T\times
SL(2,\IZ)_U\times \IZ^{T\leftrightarrow U}_2$. The field
$X^0$ accounts for the additional vector multiplet of
the graviphoton. The gauge kinetic part of the Lagrangian is
\eqn\egaugea{
\Lc_{\rm gauge}=-{i\o 4}(\Nc_{IJ}-\bar\Nc_{IJ}) F_{\mu\nu}^I F^{\mu\nu\,J}+
{1\o 4}(\Nc_{IJ}+\bar\Nc_{IJ}) F_{\mu\nu}^I \tilde F^{\mu\nu\,J}\ ,}
where the gauge couplings are then expressed in terms of ${\cal N}_{IJ}$
\eqn\proyen{
{\cal N}_{IJ}=\bar F_{IJ}+2i {{\im F_{IK} \im F_{JL} X^K X^L} \o {\im F_{MN}
X^M X^N}}  }
via
\eqn\gauge{
g^{-2}_{IJ}={\cal N}_{IJ}-{\ov {\cal N}}_{IJ}\ .}
The scalar partner (and spinor) of the graviphoton is gauge fixed 
in super Poincar{\'e} gravity. 
With the standard choice for the scalar fields of the vector multiplets
\eqn\moduli{
S={X^1 \o X^0}\ \ \ ,\ \ \ T={X^2 \o X^0}\ \ \ ,\ \ \
U={X^3 \o X^0}\ ,}
we derive from \proyen\ and \gauge\ for the effective gauge couplings
up to order $\Oc(f^2,(\p f)^2/(S-\ov S)$
\eqn\gauges{\eqalign{
g^{-2}_{22}&={(S-\bar S)(U-\bar U)\o (T-\bar
 T)}-{1\o 4}\Big[\p_TD_Tf-\p_{\bar T}D_{\bar T}\bar f\Big]-
{1\o 4}{(U-\bar U)^2\o(T-\bar T)^2}\Big[\p_U D_U f-\p_{\bar U}D_{\bar
 U}\bar f\Big]\cr
&+\h{(U-\bar U)\o(T-\bar T)}\Big[\p_T\p_U f-\p_{\bar T}
\p_{\bar U}\bar f\Big]+\Oc\lf(f^2,(\p f)^2\o S-\ov S\ri) \cr
g^{-2}_{23}&=(S-\bar S)-\h\Big[D_TD_Uf-D_{\bar T}D_{\bar U}
\bar f\Big]
+{1\o 4}{(T-\bar T)\o(U-\bar U)}\Big[\p_T D_T f-\p_{\bar T} D_{\bar T}
\bar f \Big]\cr
&+{1\o 4}{(U-\bar U)\o(T-\bar T)}\Big[\p_UD_U f-\p_{\bar U}D_{\bar
 U}\bar f\Big]+\Oc\lf(f^2,(\p f)^2\o S-\ov S\ri) \ ,}}
where the covariant derivatives are
$D_T=\p_T-{2\o (T-\bar T)}$ and $D_{\bar T}=\p_{\bar T}+{2\o (T-\bar
T)}$.
Of course, there is no dilaton dependence at one--loop. 
The expression for $g^{-2}_{33}$ we simply deduce from 
$g^{-2}_{22}$ by exchanging $T$ and $U$.
The couplings in \gauges\ are related via 
${(T-\bar T)\o (U-\bar U)}g^{-2}_{22}+g_{23}^{-2}=
2(S-\bar S)-\h\Big(D_TD_Uf-\p_T\p_U f-hc.\Big)$.

In the interpretation of \gauges\ one occurs a puzzle:
Along the explanations of above the gauge--couplings should not
receive higher than one--loop contributions, i.e. 
powers in $1/(S -\ov S)$ must not appear.
Nonetheless, it is the dilaton independent part of \gauges, which is
relevant for the string one--loop calculations in the next section.
This is in precise analogy of \afgnt, where an expansion like
in \gauges\ has been performed for the one--loop K{\"a}hler metric.
All the same, for completness, let us mention the solution to that puzzle 
in view of the gauge--couplings \proyen. The symplectic 
transformation $(X^I,F_J)\rightarrow(\hat{X}^I,\hat{F}_J)$ \dewit
\eqn\symplectic{\eqalign{
\hat{X}^1=F_1&,\ \ \hat{F}_1=-X^1\ ,\cr
\hat{X}^I=X^I&,\ \ \hat{F}_I=F_I\ , I\neq 1}}
changes the metric from ${\cal N}$ to 
$\hat{\cal N}$ with \def\n {{\cal N}}
\eqn\transmat{
\hat{\cal N}=\pmatrix{\n_{00}-{\n_{01}\n_{10}\o
\n_{11}}&{\n_{01}\o\n_{11}}&\n_{02}-{\n_{01}\n_{12}\o\n_{11}}&\n_{03}-
{\n_{01}\n_{13}\o\n_{11}}\cr\noalign{\vskip2pt}
{\n_{10}\o\n_{11}}&-{1\o\n_{11}}&{\n_{12}\o\n_{11}}&{\n_{13}\o\n_{11}}
\cr\noalign{\vskip2pt}
\n_{20}-{\n_{10}\n_{21}\o
\n_{11}}&{\n_{21}\o\n_{11}}&\n_{22}-{\n_{12}\n_{21}\o\n_{11}}&\n_{23}-
{\n_{13}\n_{21}\o\n_{11}}\cr\noalign{\vskip2pt}
\n_{30}-{\n_{10}\n_{31}\o
\n_{11}}&{\n_{31}\o\n_{11}}&\n_{32}-{\n_{12}\n_{31}\o\n_{11}}&\n_{34}-
{\n_{13}\n_{31}\o\n_{11}}\cr}\ .}
It can be verified that in this new basis \transmat, all gauge
couplings involve neither powers of $1/(S-\ov S)$ nor higher orders in
$f$ or its derivatives. This is just an effect of a rearrangement of all
couplings $\n_{IJ}$ \proyen\ in \transmat.
E.g. for ${\hat{g}^{-2}}_{22}\equiv{\hat{\n}}_{22}-
{\ov{\hat{\n}}_{22}}=
2i\im\lf(\n_{22}-{\n_{12}^2\o\n_{11}}\ri)$ one determines
\eqn\newgTT{\eqalign{
{\hat{g}^{-2}}_{22}
&=-4(\tilde S-\ov{\tilde S}) {|U|^2\o (T-\ov T)(U-\ov U)}\cr
&-{1\o(U-\ov U)^2}\lf[
\ov U^2\p_T D_T f-U^2\p_{\ov T} D_{\ov T} \ov f\ri]-{1\o(T-\ov T)^2}
\lf[U^2 \p_UD_U f-\ov U^2\p_{\ov U} D_{\ov U} \ov f\ri]\ ,\cr
{\hat{g}^{-2}}_{23}&=-(\tilde S-\ov{\tilde S}) 
{(T+\ov T)(U+\ov U)\o (T-\ov T)(U-\ov U)}-\h\lf[D_UD_Tf-D_{\ov
U}D_{\ov T}\ov f\ri]\cr
&-{1\o(U-\ov U)^2}\lf[
T\ov U\p_T D_T f-\ov T U\p_{\ov T} D_{\ov T} \ov f\ri]-
{1\o(T-\ov T)^2}\lf[\ov T U \p_U D_U f-T\ov U \p_{\ov U} D_{\ov U}\ov
f \ri]\ ,}}
with the pseudo--invariant dilaton $\tilde S$ \dewit
\eqn\psS{
\tilde S=S+\h \p_T\p_U f\ .}

\newsec{String amplitudes}
In this section we determine the one--loop correction
to the $U(1)^2_L$ gauge couplings of the effective action of
an N=2 heterotic string compactified on $K3\times T^2$ which has 
been studied in \dewit\ from a field theoretical point of view.
Here we want to focus on the derivation via string amplitudes.
To this end we calculate the CP even part of
two--point one--loop string amplitudes including gauge bosons 
of the internal Abelian gauge group of the torus $T^2$ in a background
field method.
Then we compare the $\Oc(k^2)$ piece of the string amplitudes with
the effective Lagrangian \egaugea\  of N=2 supergravity. 
We also compute three point amplitudes with two gauge bosons
and a modulus $U$ or $T$ which corresponds to derivatives of
the gauge couplings.

The relevant vertex operators in the zero ghost picture
for the moduli 
$T=T_1+iT_2=2(b+i\sqrt{G})$ and $U=(G_{12}+i\sqrt{G})/G_{11}$
w.r.t. the
background fields $G_{IJ}$ and $B_{IJ}=b\, \eps_{IJ}$ are
\eqn\vertm {
V_\pm^{(0)}= \bar\p X^\pm\,\Big[\p X^\pm+i (k\cdot \psi)\Psi^\pm\Big] 
e^{i k\cdot X}\ ,}
where $X^\pm={1\o \sqrt{2}}\Big(X_1\pm iX_2\Big)$ 
are the internal bosonic fields, $\Psi^\pm$
their supersymmetric partners and $\psi^\mu$ are spacetime fermions
with $\mu=0,\cdots,3$.
The vertex operators for the $U(1)_L^2$ gauge bosons $A_\mu^\pm$
of the internal torus
are \llt
\eqn\vertg{
V_{A_\pm}^{(0)}=\rho\ \eps_\mu\bar\p X^\pm\Big[\p X^\mu+i (k\cdot 
\psi)\psi^\mu\Big]
 e^{i k\cdot X}\ ,}
with spacetime polarization tensor $\eps_\mu$ and 
$\rho(T)=\sqrt{U_2\o T_2}$ and $\rho(U)=\sqrt{T_2\o U_2}$.

There is an important point regarding the choice of 
normalization of the vertex operators, which has two, seemingly
different, explanations: one
based on target--space--duality, i.e. string theory and one
coming from N=2 supergravity.
The stringy argument: The calculated amplitudes
have to have a certain modular weight under $T$-- and $U$--duality as
it can be anticipated from \newgTT. This
is precisely achieved by that choice.
The supergravity argument: The specific mixing between the
scalars of the vector multiplet and gauge bosons via the covariant derivative
involves a coupling which is not the gauge coupling but given by the 
K{\"a}hler metric.
If the fields and propagators are correctly normalized the 
corresponding Feynman diagram contributes only with the 
gauge coupling.

\subsec{Two-point string amplitudes}
We consider the $\Oc(k^2)$ contribution of 
the two point one--loop string amplitude  
including two gauge bosons $A_\mu^+$. It will produce a term
\eqn\stgauge
{-{i\o 4}{\Delta_{(TT)}\o 8\pi^2}F_{\mu\nu}^TF^{\mu\nu\,T}} 
in the effective action. We
denote the one--loop threshold correction to the internal $U(1)_T$ 
gauge coupling
by $\Delta_{(TT)}$. On the other hand, the 
gauge couplings \gauges, which refer to the supergravity basis, will
turn out to be linear combinations of the couplings of 
$U(1)_T$ and $U(1)_U$.

We take the gauge boson vertex operators of the two--point function
in a constant background field similarly to \vadim. 
Otherwise,  the kinematic factor  will cause the two point amplitude
to vanish. 
Thus we take 
$A_\mu^+(X)=-\h F_{\mu\nu}^T X^\nu$ with $F_{\mu\nu}^T={\rm const}$ and
the polarization tensor of the gauge boson $A_\mu^+$ is 
replaced by  $\eps_\mu e^{i k\cdot X}\to A^+_\mu(X)$. The vertex
operator of the gauge bosons is then
$\tilde V^{(0)}_{A_+}=-\h F_{\mu\nu}^T\rho(T)\bar\p X^+(X^\nu \p X^\mu+
\psi^\mu\psi^\nu)$.
The general  expression for the CP even part of the
string amplitude is \DKLII 
\eqn\eampl{\Ac(A_+,A_+)=\sum_{\rm even\, s}(-1)^{s_1+s_2}\int_{\Gamma} d^2\tau
\,Z(\tau,\bar\tau,s)\int d^2z_1\la
\tilde V_{A_+}^{(0)}(z,\bar z)\tilde V_{A_+}^{(0)}(0)\ra\ ,}
where 
\eqn\part{
Z(\tau,\bar\tau,s)={\rm Tr}_{s_1}\lf[(-1)^{s_2 F} q^{H-\h}\bar
q^{\bar H-1}\ri]=Z_\psi Z_X Z_{X_0} Z_{\rm int} }
is the partition function ($q=e^{2\pi i\tau},\ov q=e^{-2\pi i\ov\tau}$)
for even spin structures $(s_1,s_2)=(1,0),(0,0),(0,1)$ and
$Z_\psi={\th_\alpha(0,\tau)\o \eta(\tau)}$ 
the fermionic partition function  where $\th_\alpha$
are the Riemann theta functions for $\alpha=2,3,4$.
The contribution
from bosonic zero modes is $Z_{X_0}={1\o 32 \pi^4 \tau_2^2}$
and $Z_X={1\o |\eta(\tau)|^4}$ is the bosonic partition function.
The fermion number is denoted by $F$.
The integration region is the fundamental region of the worldsheet torus
$\Gamma=\{\tau :\ |\tau_1|\le\h,|\tau|\ge 1\}$.

After summing over even spin structures
we only get non vanishing contributions 
if four space-time fermions are contracted because of 
a theta function identity. Therefore, pure bosonic contractions may be omitted.
The two point function gives\foot{We
introduced the notation $X_i=X(z_i,\bar z_i)$.}
\eqn\tt
{\la \tilde V_{A_+}^{(0)}(\bar z,z)\tilde V_{A_+}^{(0)}(0)\ra=-\h
F_{\mu\nu}^TF^{\mu\nu\,T}\rho(T)^2 G_F^2\la\bar \p X_1^+\bar
\p X^+_2\ra\ ,}
where 
$G_F={\th_1(0,\tau)\th_\alpha(z,\tau)\o \th_\alpha(0,\tau)\th_1(z,\tau)}$
with $\alpha=2,3,4$ is
the fermionic Green function and the part of
$G_F^2$ depending on spin structures is $4\pi i\p_{\tau}\ln Z_\psi$
which does no longer depend on worldsheet coordinates.
$G_B=-\ln|\chi|^2$ is the
bosonic Green function with $|\chi|^2=4\pi^2 e^{-2\pi (\im
\,z)^2/\im\tau}\Big|{\th_1(z,\tau)\o\th_1(0,\tau)}\Big|^2$. 
We take the following Green functions for the internal bosons 
\eqn\ig{
\eqalign{
\la\bar\p X^\pm\bar\p X^\pm\ra&=2\pi^2(P_R^\pm)^2\cr
\la\bar\p X^\pm\bar\p X^\mp\ra &=2\pi^2 P_R^\pm P_R^\mp-{\pi\o\tau_2}
+\bar \p^2 G_B\cr
\la\bar\p X^\pm\p X^\mp\ra &=2\pi^2 P_R^\pm P_L^\mp-\pi \delta^{(2)}(z)
\ ,}}
with Narain momenta $P_{R/L}^+=\bar P_{R/L}$ and $P_{R/L}^-=P_{R/L}$
which are defined as
\eqn\momenta{
\eqalign{
P_L&={1\o \sqrt{2T_2 U_2}}\Big(m_1+m_2\bar U+n_1\bar T+n_2\bar T\bar U\Big)\cr
P_R&={1\o \sqrt{2T_2 U_2}}\Big(m_1+m_2\bar U+n_1 T+n_2 T\bar U\Big)\ .}}
>From $\int d^2z \,\p^2 G_B=0$ we get
$\int d^2 z\, \p^2 \ln\th_1(z,\tau)=-\pi$. Using this relation
we find the following result
\eqn\att{
\Ac(A_+,A_+)=-{i\o 4} F_{\mu\nu}^TF^{\mu\nu\,T}\Delta_{(TT)}\ ,}
with
\eqn\dtt{
\Delta_{(TT)}=-{U_2\o T_2}\int 
{d^2\tau\o\tau_2} 
\bigg[\sum_{P_L,P_R} q^{\h |P_L|^2}\bar q^{\h |P_R|^2}
\bar P_R^2\bigg]\bar F_{-2}(\bar\tau)\ ,}
where $\ov F_{-2}={\ov E_4\ov E_6\o \ov\eta^{24}}$ and
we define
\eqn\torus{
Z_{torus}(\tau,\ov\tau)=\sum_{(P_L,P_R)}q^{\h |P_L|^2}
\ov q^{\h |P_R|^2}:=\sum_{(P_L,P_R)} \hat{Z}_{torus}\ .}

The string amplitude involving
$\la \tilde V^{(0)}_{A_-}\tilde V^{(0)}_{A_-}\ra$ is easily 
obtained from \att\ 
by exchanging $T_2$ with $U_2$ and replacing $\bar P_R$ with its
complex conjugate $P_R$.
Similarly, for the string amplitude $\la
\tilde V^{(0)}_{A_+} \tilde V^{(0)}_{A_-}\ra$ we get the modular invariant 
result: 
\eqn\atu{
\Ac(A_+,A_-)={i\o 4} F_{\mu\nu}^TF^{\mu\nu\,U}\int{d^2\tau\o
\tau_2}\sum_{(P_L,P_R)}
\Big(|P_R|^2-{1\o 2\pi\tau_2}\Big)
\hat{Z}_{torus}\bar F_{-2}(\bar\tau)\ .}
This result can be directly compared with the one loop correction to the
K{\"a}hler potential 
$G_{T\bar T}^{(1)}=-{i\o 2}G_{T\bar T}^{(0)} D_T D_U f+hc.$ which has been
derived in \afgnt.
The second part may be identified with the Green-Schwarz (GS)--term 
$2 G^{(1)}\equiv\Delta_{\rm univ}=\int {d^2\tau\o \tau_2} \Big(-{1\o
2\pi\tau_2}\Big) Z_{torus}\bar F_{-2}$.

Our results \att\ and \atu\ refer to the string basis (3.1) and (3.2) and
(therefore) involve modular invariant integrands.
Since the momenta transform under $SL(2,{\bf Z})_T\times SL(2,{\bf
Z})_U$, like $(ad-bc=1)$
\eqn\dulnar{\eqalign{
(P_L,\bar P_R)&\to\sqrt{cT+d \o c\bar T+d}\ (P_L,\bar P_R)\ \ \ ,\ \ \
T\to {{aT+b}\o {c T+d}}\ , \ U\to U\ , \cr
(P_L,P_R)&\to\sqrt{cU+d\o c\bar U+d}\ (P_L,P_R)\ \ \ ,\ \ \ 
T\to T\ ,\ U\to {{a U+b} \o {c U+d}\ }\ ,}}
we realize that these amplitudes transform with specific weights
$(w_T,w_U)=(2,-2)$, $(w_{\bar T},w_{\bar U})=(0,0)$ and 
$(w_T,w_U)=(0,0)$,  $(w_{\bar T},w_{\bar U})=(0,0)$, respectively.
They can be directly
identified with well--defined integrals $\Ic_{2,-2}$ and $\Ic_{0,0}$
as will be shown in the next section. 

The one--loop correction to the gauge coupling $g_{22}^{-2}$, as it
has appeared in the last section and which is therefore w.r.t. 
the supergravity basis, is then obtained 
by taking a linear combination of string amplitudes \att\ and \atu,
which corresponds
to the correlation function of two  gauge boson vertex operators  
$A_\mu^{T}=\h[A_{\mu}^+-{(U-\bar U)\o (T-\bar T)}A_\mu^-]$:
\eqn\gloop
{\eqalign{
\Big[g_{22}^{-2}\Big]^{1-loop}&={1\o 4}
\bigg[{\Delta_{(TT)}\o 8\pi^2}-
2{(U-\bar U)\o(T-\bar T)}{\Delta_{(TU)}\o 8\pi^2}
+{(U-\bar U)^2\o (T-\bar T)^2}
{\Delta_{(UU)}\o 8\pi^2}\bigg]-{1\o 8\pi^2}{U_2\o T_2}G^{(1)}\cr
&=-{1\o 32\pi^2}\bigg[{U-\bar U\o T-\bar T}\int {d^2\tau\o\tau_2}
\sum_{(P_L,P_R)} (\bar P_R-P_R)^2 \hat Z_{torus}\bar
F_{-2}(\bar\tau)\bigg]\ .}}
Notify, that in the above expression 
the GS-term cancels the one in $\Delta_{(TU)}$.
After symplectic transformation \symplectic\ to the gauge coupling 
${\hat g}_{22}^{-2}$ one obtains for its loop--correction:
\eqn\ghloop
{\eqalign
{\Big[{\hat{g}}_{TT}^{-2}\Big]^{1-loop}&=
{{\bar U}^2\o (U-\bar U)^2}{\Delta_{(TT)}\o 8\pi^2}
+{U^2\o (T-\bar T)^2}
{\Delta_{(UU)}\o 8\pi^2}+2{|U|^2\o (U-\bar U)(T-\bar T)}
{\Delta_{(TU)}\o 8\pi^2}\cr
&+{|U|^2\o (T-\bar T)(U-\bar U)}
{G^{(1)}\o 2\pi^2}\cr
&=-{1\o 8\pi^2}\bigg[{1\o (T-\bar T)(U-\bar U)}\int {d^2\tau\o\tau_2}
\sum_{(P_L,P_R)} (\bar U\bar P_R+U P_R)^2 \hat Z_{torus}\bar
F_{-2}(\bar\tau)\bigg]\ .}}
This amplitude can be directly derived from a two point amplitude with
vertex operators corresponding to 
$\hat{A}_\mu^T={\bar U\o(U-\bar U)}A_\mu^+-{U\o (T-\bar T)}A_\mu^-$.
Using the results from the next section we may directly cast
\gloop\ and \ghloop\ into the forms (2.6) and (2.9), dictated by
supergravity.

In the corrections \gloop\ and \ghloop, there appears the non--modular
invariant GS--term $G^{(1)}$. On the other hand, the dilaton field $S$
gets modified at one--loop by the same amount with an opposite sign \dfkz.
Thus, altogether, the physical coupling stays modular invariant.
See also \nsii\ for a more complete discussion. 

\subsec{Three-point amplitudes}
Now we consider three point amplitudes which involve two internal
gauge bosons and one modulus. 
First we want to investigate the amplitude including two gauge bosons
$A_T$ and one $T$ modulus 
which is related to $ \p_T\Delta_{(TT)} F_{\mu\nu}F^{\mu\nu}T$ in the
effective string action, where $\p_T\Delta_{(TT)}$ denotes
the derivative with respect to $T$ of the one loop correction to the 
$U(1)_T$ gauge coupling. 
The correlation function gives the following contractions:
\eqn\ttt
{\eqalign{\la V_+^{(0)}(z_1)V_{A_+}^{(0)}(z_2)V_{A_+}^{(0)}(z_3)\ra&=\Kc
\,G_F^2\prod_{i<j}|\chi_{ij}|^{2 k_i\cdot k_j} {U_2\o i T_2^2} \cr
&\Big(\la\bar\p X_1^+\p X_1^+\ra\la\bar \p X_2^+\bar \p X_3^+\ra
+\la\bar\p X_1^+\bar \p X_2^+\ra\la\p X_1^+\bar \p X_3^+\ra\cr
&+\la\bar\p X_1^+\bar
\p X_3^+\ra\la\p X_1^+\bar\p X_2^+\ra\Big) ,}}
where $\Kc=\Big((k_2k_3)(\eps_2\eps_3)-(k_2\eps_3)(k_3\eps_2)\Big)$ is
the kinematic factor.

Before doing the worldsheet integrals we want to make some comments
on possible additional non trivial contributions to the $\Oc(k^2)$ part of
the amplitude. 
We may get contributions from the
delta function which might appear in the correlation function $\la \bar
\p X^\pm\p X^\mp\ra$. If we consider the region
$|z_{ij}|<\eps$ then $|\chi_{ij}|\simeq |z_{ij}|$ and thus the delta function
can be omitted because  $\int d^2z_i\,
\delta^{(2)}(z_{ij}) |z_{ij}|^{2 k_i\cdot k_j} f(z_{ik})=0$ where
$f$ is some function. 
But if $|z_{ij}|> \eps$ and $|k_i\cdot k_j G_{ij}^B|\ll 1$ then
one can expand $|\chi_{ij}|^{2 k_i\cdot k_j}=1-k_i\cdot k_j
G_{ij}^B+\ldots$  and in this case one indeed gets contributions
from the delta function for the lowest term of the expansion
\ejm.
On the other hand, if the correlation functions can be approximated such
that the worldsheet integral gives
$\int_{|z_{il}|<\eps} d^2 z_{il} {|z_{il}|^{2
k_i\cdot k_l}\o |z_{il}|^2}\sim {\pi\o k_i\cdot k_l}$ 
one may e.g. produce a $\Oc(k^2)$ contribution from  terms of the
order $\Oc(k^4)$. These
contributions are important when one has to collect 
all possible terms of a particular order \min.
But in the case considered here,  pinched off integrals 
only give $\int_{|z_{il}|<\eps} d^2 z_{il} {|z_{il}|^{2
k_i\cdot k_l}\o |z_{il}|^4}\simeq  {\pi\o k_i\cdot k_l-1}$ and thus do
not contribute to the $\Oc(k^2)$ piece of the amplitude. 

In the following we will restrict ourselves to the region
$|z_{ij}|<\eps$. Taking into account the arguments mentioned above 
it remains to perform the worldsheet integral of  \ttt. 
We end up with:
\eqn\attt{
\Ac(T,A_+,A_+)|_{\Oc(k^2)}=-\Kc{\pi^2 \ U_2\o 2\  T_2^2}\int
d^2\tau
\bigg\{\sum_{P_L,P_R} q^{\h |P_L|^2}\bar q^{\h |P_R|^2} \bar P_R^3
P_L \bigg\} 
\bar F_{-2}(\bar\tau)\,}
This term can be identified with the third
derivative of the prepotential\foot{The relevant relations between the 
prepotential and $\tau$--integrals may be found in the next section.} 
$f_{TTT}$ which  has been derived in \afgnt \ by taking particular
derivatives on the integral coming from a CP odd  string amplitude
of the one loop correction to the K{\"a}hler potential
$G_{T\bar T}^{(1)}$. 
Thus one finds
\eqn\fattt{
\Ac(T,A_T,A_T)|_{\Oc(k^2)}=-4 i\Kc\pi^3 f_{TTT}\ .}
We will have to say more about this result in section 4.
We realize that this expression transforms covariantly 
under $SL(2,{\bf Z})_T\times SL(2,{\bf Z})_U$
with weights $(w_T=4,w_{\bar T}=0)$
and $(w_U=-2,w_{\bar U}=0)$, respectively.

Besides we calculate the three point amplitudes 
$\la V_+^{(0)}V^{(0)}_{A_+}V_{A_-}^{(0)}\ra$ and
$\la V_+^{(0)}V^{(0)}_{A_-}V_{A_-}^{(0)}\ra$
with the result
\eqn\attu
{\eqalign
{\Ac(TA_+ A_-)&=-{\Kc\pi^2\o 2\ T_2}\int d^2\tau
\bigg\{\sum_{P_L,P_R} q^{\h |P_L|^2}\bar q^{\h |P_R|^2}
\Big[ \bar P_R |P_R|^2 P_L-{1 \o \pi\tau_2}\bar P_R P_L\Big]\bigg\} 
\bar F_{-2}(\bar\tau)\cr
\Ac(TA_- A_-)&=-{\Kc\pi^2\o 2\ U_2}
\int d^2\tau
\bigg\{\sum_{P_L,P_R} q^{\h |P_L|^2}\bar q^{\h |P_R|^2}
\Big[P_R |P_R|^2 P_L-{1\o \pi\tau_2}P_RP_L
\Big]\bigg\} \bar F_{-2}(\bar\tau)\ .}}
Similarly to \fattt \ these amplitudes can be casted into the 
convenient form as it will be shown in section 4.
\eqn\ampli{\eqalign{
\Ac(TA_+A_-)&=-4i\Kc\pi^3
\bigg[ f_{TTU}+ {1\o 4\pi^2}G_T^{(1)}\bigg]\cr
\Ac(TA_-A_-)&=-4i\Kc\pi^3
\bigg[ f_{TUU}+ {1\o 4\pi^2}G_U^{(1)}\bigg]\ .}}

The linear combination which corresponds to the three point amplitude
with the $T$ modulus  and two $A_\mu^{T}$ gauge bosons is
\eqn\lincomb{\kern-1em\eqalign{
\Ac(T A^{T}A^{T})&={1\o 4}\bigg[\Ac(TA_+A_+)-2{(U-\bar U)\o (T-\bar T)}
\Ac(TA_+A_-)+{(U-\bar U)^2\o (T-\bar T)^2}\Ac(TA_-A_-)\bigg]\cr
&=-{\Kc\pi^2\o 8}{U_2\o T_2^2}\int d^2\tau
\bigg\{\sum_{P_L,P_R} q^{\h |P_L|^2}\bar q^{\h |P_R|^2}
P_L \bar P_R(\bar P_R-P_R)^2\cr
&+{2\o \pi\tau_2}P_L\bar P_R-{1\o\pi\tau_2}P_LP_R
\bigg\} \bar F_{-2}(\bar\tau)\ .}}
The string amplitude is not a 1PI diagram but 
also contains other exchange diagrams and 
therefore splits into field theoretical amplitudes containing 
one loop corrections to the gauge coupling.
\eqn\ft{\kern-1em
\Ac(T A^{T}A^{T})={\Kc\pi U_2\o 4 i T_2}\lf[16\pi^2\p_T g_{23}^{-2}
 -{1\o 4 i U_2}\Delta_{(TT)}+{U_2\o 4 i T_2^2}\Delta_{(UU)}\ri]\ .}

\newsec{Prepotential and world--sheet torus integrations}

In this section we want to find relations of the one--loop
prepotential $f$ and/or derivatives of it to world--sheet
$\tau$--integrals as they appear in the previous section.
The one--loop correction\foot{Compared to the 
previous sections
we now change $f\to i f$.} to the heterotic prepotential \prep\
can be written in the chamber $T_2>U_2$ \hm

\eqn\fone{
f(T,U)={i\o 12\pi }U^3
+{1\o (2\pi i)^4}\sum_{(k,l)>0} c_1(kl)\ \Li_3 \lf[e^{2\pi
i(kT+lU)}\ri] 
+{1\o 2(2\pi)^4} c_1(0)\zeta(3)\ ,}
where the numbers $c_1(n)$ are related to the (new) supersymmetric index 

\eqn\susyindex{
{\cal Z}(\tau,\ov\tau)=\ov \eta^{-2}\ {\rm Tr}_R\lf[F(-1)^F\ 
q^{L_0-{c\o 24}}\ov q^{\ov L_0-{\ov c\o 24}}\ri]\ ,}
which for heterotic compactifications on $K3\times T^2$ with 
the choice of $SU(2)$ instanton numbers 
$(12,12),\ (11,12)\ ,(10,14)$ and $(24,0)$ becomes \hm
\eqn\index{\eqalign{
{\cal Z}(\tau,\ov\tau)&=2i\ Z_{torus}(\tau,\ov\tau)
{\bar E_4 \bar E_6 \o \ov \eta^{24}}\ ,\cr
{\bar E_4 \bar E_6 \o \ov \eta^{24}}&=\sum_{n\geq -1} c_1(n)\ \bar
q^n\ .}}
The mentioned models lead to the
gauge group $E_7\times E_7$ and $E_7\times E_8$, respectively.
In the first three cases the gauge group may be completely Higgsed away.
At the perturbative level these models are equivalent. A fact, which
also becomes clear from the unique expression for the supersymmetric 
index \index\ which enters all kinds of perturbative string
calculations (cf. e.g. the previous section). These three models
(after Higgsing completely) are 
dual to typeIIA Calabi--Yau compactifications, which are elliptic 
fibrations over the Hirzebruch surfaces $\IF_0,\IF_1,\IF_2$. 
Then the holomorphic part (to be identified with the Wilsonian
coupling) of the three-point functions \attt\ and \attu\ 
[in particular (3.18) and (3.20)] is related to the Yukawa couplings 
$f_{TTT}, f_{TTU}$ and $f_{TUU}$ 
of the Calabi--Yau manifold, respectively \klm\klt. 
Moreover using mirror symmetry
these couplings are given by the classical intersection numbers of the
typeIIB theory.
Supersymmetric indices \susyindex, valid for the other
bases $\IF_k$ are the subject of \st. They allow for more general
instanton embeddings and one ends up with larger terminal gauge groups after
Higgsing.

We introduce the polylogarithms $(a \geq 1)$:
\eqn\LIS{
\Li_a(x)=\sum_{p>0} {x^p \o p^a}\ .}

The integrals we should look for involve Narain momenta from `charge' 
insertions or zero--mode contributions.
In general they show up
in string amplitudes involving $U(1)$--charges w.r.t. the internal 
bosonic fields or after contractions of bosonic internal 
fields (belonging to the $T^2$).
I.e. we consider  ($\alpha,\beta,\gamma,\delta \geq 0$)
\eqn\setup{
{\cal I}_{w_T,w_U}:=
(T- \ov T)^m (U-\ov U)^n\ \int {d^2\tau \o \tau_2^k}\  \sum_{(P_L,P_R)}
P_L^\alpha P_R^\beta \ov P_L^\gamma \ov P_R^\delta\   
{\hat Z}_{torus}(\tau,\ov\tau)  \ \ov F_{l}(\ov \tau)\ .} 
We want this expression to have modular weights $(w_T,w_U)$
and weights $(w_{\ov T},w_{\ov U})=(0,0)$ under $T,U$--duality.
This imposes the conditions [cf. \dulnar]:

\eqn\conds{\eqalign{
m&=-{w_T \o 2}\cr
n&=-{w_U \o 2}\cr
\gamma&=\alpha-\h(w_T+w_U)\cr
\delta&=\beta +\h(w_T-w_U)\ .}}
There is also a relation for $k$ and $l$ which follows from modular
invariance of the integrand, which can be easily deduced after a
Poisson resummation on the momenta $m_i$.
Since the integrals \setup\ will be constructed such that they 
transform with a certain weight under $SL(2,\IZ)_T\times SL(2,\IZ)_U$
we expect that ${\cal I}_{w_T,w_U}$ can be written in terms of modular
covariant derivatives of $f$ rather than usual derivatives.
The prepotential $f(T,U)$ has weights $(w_T,w_U)=(-2,-2)$.
Acting with the covariant derivative (cf. also section 2)
\eqn\covT{
D_T=\p_T-{2\o T-\ov T}}
increases its weight $w_T$ by $2$.
In general with the derivative
\eqn\covsT{
D^n_T=\p_T-{2n\o T-\ov T}}
one changes the weight from $-2n$ to $-2n+2$.
This derivative is also covariant w.r.t.
the K{\"a}hler connection $a_\mu \sim [\p_i K(\Phi,\bar\Phi)D_\mu\phi^i-
\p_{\bar i} K(\Phi,\bar\Phi)D_\mu \bar \phi^{\bar i}]$, which means
from the point of view of amplitudes that one--particle reducible
diagrams with massless states running in the loop are subtracted 
to end up with the 1PI effective action.

In subsection 4.1 we consider cases involving only two momenta, i.e.
$\alpha+\beta+\gamma+\delta=2$. 
In subsection 4.2. some cases of more than two momenta insertions.

\subsec{Two momenta insertions}
\subsubsec{4.1.1.\  $f_{TT}$}

Let us consider the integral

\eqn\ftt{
{\cal I}_{2,-2}:=
{(U- \ov U)\o  (T-\ov T)}\ \int {d^2\tau \o \tau_2}\  \sum_{(P_L,P_R)}
\ov P_R^2\   
{\hat Z}_{torus}(\tau,\ov\tau)\ {\ov F}_{-2}(\ov \tau)\ ,} 
which after a Poisson resummation on $m_i$ (cf. Appendix A) becomes

\eqn\ftts{
{\cal I}_{2,-2}=
{(U- \ov U)\o  (T-\ov T)}\ T_2^2
\int {d^2\tau \o \tau_2^4}\ \sum_{n_1,n_2 \atop  l_1,l_2}
\ov Q_R^2\  e^{-2\pi i \ov T \det A}  e^{{-{\pi T_2} \o {\tau_2
U_2}}\lf|n_1\tau+n_2U\tau -Ul_1+l_2 \ri|^2} 
 \ {\ov F}_{-2}(\ov \tau)\ .} 
with $A=\lf( {n_1 \atop n_2}{-l_2\atop l_1}\ri)\in M(2,\IZ)$ and we introduce:
\eqn\Qs{\eqalign{
Q_R&={1 \o \sqrt{2T_2U_2}}\lf[(n_2 \ov U+n_1)\tau-\ov U l_1+l_2\ri] \cr
\ov Q_R&={1 \o \sqrt{2T_2U_2}}\lf[(n_2 U+n_1)\tau-U l_1+l_2\ri] \cr
Q_L&={1 \o \sqrt{2T_2U_2}}\lf[(n_2 \ov U +n_1)\ov\tau-\ov U l_1+l_2\ri]\cr
\ov Q_L&={1 \o \sqrt{2T_2U_2}}\lf[(n_2 U+n_1)\ov\tau-U l_1+l_2\ri]\ .}} 
In that form \ftts\ one easily checks modular invariance, i.e.
one deduces the only possible choice for $k$ and $l$ in \setup.
For the anti--holomorphic function we choose
\eqn\Fanti{
\ov F_{-2}(\ov \tau)={\bar E_4 \bar E_6 \o \bar\eta^{24}}} 
as it arises in physical amplitudes (cf. e.g. section 3).
Modular invariance also enables us to use the orbit decomposition 
used in \DKLII, i.e. decomposing the set of all
matrices $A$ into orbits of $SL(2,\IZ)$:
\eqn\orbits{\eqalign{
I_0&:\ A=\lf( {0 \atop 0}\ {0\atop 0}\ri)\ , \cr
I_1&:\ A=\pm\lf( {k \atop 0}\ {j\atop p}\ri)\ \ ,\ \ 0\leq j<k\ ,\
p\neq 0\ , \cr
I_2&:\ A=\lf( {0 \atop 0}\ {j\atop p}\ri)\ \ ,\ \ (j,p) \neq
(0,0)\ .}}
Clearly, $I_0$ does not give any contribution.
The remaining $\tau$--integrals $I_1$
and $I_2$ are presented in appendix B and we evaluate
for \ftt:
\eqn\fttfinal{
{\cal I}_{2,-2}=8\pi^2 \p_T\lf(\p_T-{2\o T-\ov T}\ri)f+8\pi^2
{(U-\ov U)^2\o (T-\ov T)^2} \p_{\ov U}\lf(\p_{\ov U}+{2\o U-\ov
U}\ri)\ov f\ .}
It is quite remarkable, how e.g. the cubic term of the prepotential
\fone, in the combination of \fttfinal, gives the last term of (B.19).


\subsubsec{4.1.2\ $f_{UU}$}

Similary, an expression with modular weights $(w_T,w_U)=(-2,2)$ can be
found:
\eqn\fuu{
{\cal I}_{-2,2}:=
{(T- \ov T)\o  (U-\ov U)}\ \int {d^2\tau \o \tau_2}\  \sum_{(P_L,P_R)}
P_R^2\   
{\hat Z}_{torus}(\tau,\ov\tau)\ {\ov F}_{-2}(\ov \tau)\ ,} 
which after a Poisson resummation on $m_i$ (cf. Appendix) becomes
\eqn\ftts{
{\cal I}_{-2,2}=
{(T- \ov T)\o  (U-\ov U)}T_2^2
\int {d^2\tau \o \tau_2^4}\ \sum_{n_1,n_2 \atop  l_1,l_2}
Q_R^2\  e^{-2\pi i \ov T \det A}  e^{{-{\pi T_2} \o {\tau_2
U_2}}\lf|n_1\tau+n_2U\tau -Ul_1+l_2 \ri|^2} 
 \ {\ov F}_{-2}(\ov \tau)\ .}
After the integration we end up with:
\eqn\fuufinal{
{\cal I}_{-2,2}=8\pi^2 \p_U\lf(\p_U-{2\o U-\ov U}\ri)f+8\pi^2
{(T-\ov T)^2\o (U-\ov U)^2} \p_{\ov T}\lf(\p_{\ov T}+{2\o T-\ov
T}\ri)\ov f\ .}
Alternatively, with mirror symmetry $T\leftrightarrow U$, which induces
the action $P_L\leftrightarrow P_L, P_R\leftrightarrow \ov P_R$ 
on the Narain momenta one may obtain \fuufinal\ from \ftt.

\subsubsec{4.1.3\ $f_{TU}$}

There are several ways to construct from \setup\ $\tau$--integrals
of weights $w_T=0,w_U=0$ which involve at most two Narain momenta insertions.
Let us take 
\eqn\ftuI{
{\cal I}^a_{0,0}:=
\int {d^2\tau \o \tau_2}\  \sum_{(P_L,P_R)}
{\hat Z}_{torus}(\tau,\ov\tau)\ov F_0(\tau,\ov\tau)\ \ ,} 
with ($\hat{E}_2=E_2-{3\o \pi\tau_2}$)
\eqn\Fzero{
\ov F_0(\tau,\ov\tau)=\lambda_1  
{\hat{\bar E}_2\bar E_4\bar E_6\o \bar\eta^{24}}+
\lambda_2{\bar E_6^2\o \bar \eta^{24}}
+\lambda_3{\bar E_4^3\o \bar \eta^{24}}\ }
and we choose $-264\lambda_1-984\lambda_2+744\lambda_3=0$ to avoid
holomorphic anomalies arising from triangle graphs involving two gauge
fields and the K{\"a}hler-- or sigma model connection 
as external legs with massless states running in the loop. In other
words, we want to discard non--harmonic $\ln(T-\ov T)(U-\ov U)$ terms.
Later we will see that this combination is precisely related to the 
`physical' choice \Fanti\ for
$(\lambda_1,\lambda_2,\lambda_3)=(1/6,1/3,1/2)$. In fact:
\eqn\diff{
{1\o 2\pi i}{\p \o \p\ov\tau}\ov F_{-2}(\ov\tau)=
{1\o 6}{\bar E_2\bar E_4\bar E_6\o \eta^{24}}+
{1\o 3}{\bar E_6^2\o \bar\eta^{24}}
+{1\o 2}{\bar E_4^3\o \bar \eta^{24}}\ .}
The expression ${\cal I}^a_{0,0}$ gives a 
representation for a weight zero automorphic form. However,
non--harmonic, because of $\hat{E}_2$ in \diff. Therefore the theorem
of Borcherds \borch\ does not apply. 
Using results of \hm\ it easily can be
evaluated\foot{In \hm\ the N=2 Green--Schwarz term $G^{(1)}$ is
denoted by $\triangle_{\rm univ}$.}:
\eqn\ftuIres{
{\cal I}^a_{0,0}=16\pi^2 \re(f_{TU})+2G^{(1)}\ .}
Using the explicit form of the N=2 version of the GS--term \hm
\eqn\GS{
G^{(1)}={32\pi^2 \o  (T-\ov T)(U-\ov U)} 
\re\lf\{f-\h(T-\ov T)\p_Tf-\h (U-\ov U) \p_U f\ri\}\ ,}
we arrive at

\eqn\final{
\re\lf\{ D_T D_U f \ri\}={1 \o 16 \pi^2} \int 
{d^2\tau \o \tau_2}\  \sum_{(P_L,P_R)} {\hat Z}_{torus}(\tau,\ov\tau)\ \lf[
{1\o 6}{\hat{\bar E}_2\bar E_4\bar E_6\o \bar\eta^{24}}+{1\o 3}
{\bar E_6^2\o \bar \eta^{24}}+{1\o 2}{\bar E_4^3\o \bar
\eta^{24}}\ri]\ .}

In addition, we want to investigate the integral
\eqn\ftuII{
{\cal I}^b_{0,0}:=
\int {d^2\tau \o \tau_2}\  \sum_{(P_L,P_R)}
\lf(|P_R|^2-{1 \o 2 \pi \tau_2}\ri)\   
{\hat Z}_{torus}(\tau,\ov\tau)\ {\ov F}_{-2}(\ov \tau)\ ,}
The additional GS--like term is needed to
guarantee modular invariance. That can be seen after performing
the Poisson resummation, which yields:

\eqn\ftus{
{\cal I}^b_{0,0}=-T_2^2
\int {d^2\tau \o \tau_2^4}\ \sum_{n_1,n_2 \atop  l_1,l_2}
Q_R \ov Q_R\  e^{-2\pi i \ov T \det A}  e^{{-{\pi T_2} \o {\tau_2
U_2}}\lf|n_1\tau+n_2U\tau -Ul_1+l_2 \ri|^2} 
 \ {\ov F}_{-2}(\ov \tau)\ .}
Again, this integral can be determined using formulas of the appendix
B with the result:  
\eqn\ident{
{\cal I}^a_{0,0}={\cal I}^b_{0,0}\ .}

Let us now come to the identity \diff, which is the link to 
\ident. Rewriting ${\cal I}^b_{0,0}$ as
\eqn\tauder{
{\cal I}^b_{0,0}={i \o \pi}\int {d^2\tau \o \tau_2^2}\ \p_{\ov \tau}(\tau_2 
{Z}_{torus})\ {\ov F}_{-2}(\ov\tau)}
and using \diff\ we may also deduce \ident\ after partial integration.


\subsec{More than two momenta insertions}
Let us give some representative examples.

\subsubsec{4.2.1.\ $f_{TTT}$}

We want to consider the integral
\eqn\fttt{
{\cal I}_{4,-2}:=
{(U- \ov U) \o  (T-\ov T)^2}\ \int d^2\tau
\ \sum_{(P_L,P_R)} P_L \ov P_R^3\ {\hat Z}_{torus}(\tau,\ov\tau)
\ {\ov F}_{-2}(\ov \tau)\ .}
With the identity
\eqn\ida{
P_L\ov P_R^3{\hat Z}_{torus}={{T-\ov T}\o{2\pi \tau_2}}\ \ov P_R^2\ \p_T {\hat 
Z}_{torus}}
we are able to `transform' \fttt\ into a two--momentum integral
of the kind we have discussed before. In particular, this identity tells us
\eqn\covid{
2\pi {\cal I}_{4,-2}=\lf(\p_T+{2\o T-\ov T}\ri){\cal I}_{2,-2}\ .}
Using \fttfinal\ we
obtain after some straightforward algebraic manipulations:
\eqn\ftttfinal{  
{\cal I}_{4,-2}=4\pi f_{TTT}\ .}
This identity was already derived in \afgnt, however in a 
quite different manner.
Moreover, we also may directly integrate \fttt\ as we have done so in the
last subsections.
After a Poisson resummation [cf. (A.1) and 
(A.5)] the integral \fttt\ becomes : 
\eqn\FTTT{\eqalign{
{\cal I}_{4,-2}&=
-{(U- \ov U) \o  (T-\ov T)^2}\ T_2^4\ \int {d^2\tau \o \tau_2^5}\ 
\sum_{n_1,n_2 \atop  l_1,l_2} Q_L\ov Q_R^3\  e^{-2\pi i \ov T \det A}  
e^{{-{\pi T_2} \o {\tau_2 U_2}}\lf|n_1\tau+n_2U\tau -Ul_1+l_2 \ri|^2} 
\ {\ov F}_{-2}(\ov \tau)\cr
&+{3 \o 2\pi}\ {(U- \ov U) \o  (T-\ov T)^2}\ T_2^2 
\int {d^2\tau \o \tau_2^4}\ 
\sum_{n_1,n_2 \atop  l_1,l_2} 
\ov Q_R^2\  e^{-2\pi i \ov T \det A}  e^{{-{\pi T_2} \o {\tau_2 U_2}}
\lf|n_1\tau+n_2U\tau -Ul_1+l_2 \ri|^2} 
 \ {\ov F}_{-2}(\ov \tau)\ .}}
The second integral is of the kind \ftts. 
In fact, using the results of appendix B, we have explicitly
evaluated the integrals 
\FTTT\ and checked \ftttfinal.

\subsubsec{4.2.2.\ $f_{TTU}$}

A covariant expression ${\cal I}_{2,0}$ containg $f_{TTU}$ may be found
by considering $\p_T {\cal I}_{0,0}$:
\eqn\fttu{
{\cal I}_{2,0}:=\p_T {\cal I}^b_{0,0}=
{2\pi\o  (T-\ov T)}\ \int d^2\tau
\  \sum_{(P_L,P_R)} \lf(P_L P_R \ov P_R^2-{P_L\ov P_R\o \pi\tau_2}\ri)   
{\hat Z}_{torus}(\tau,\ov\tau)\ {\ov F}_{-2}(\ov \tau)\ .}
Whereas in \fttu\ each term alone already has weights $(w_T,w_U)=(2,0)$  
and $(w_{\ov T}, w_{\ov U})=(0,0)$, only their combination gives rise 
to a modular invariant integrand. This may be seen after doing the Poisson 
transformation:
\eqn\FTTU{\eqalign{
{\cal I}_{2,0}&=
{2\pi \o  (T-\ov T)}\ T_2^4\ \int {d^2\tau \o \tau_2^5}\ 
\sum_{n_1,n_2 \atop  l_1,l_2} Q_LQ_R\ov Q_R^2\  e^{-2\pi i \ov T \det A}  
e^{{-{\pi T_2} \o {\tau_2 U_2}}\lf|n_1\tau+n_2U\tau -Ul_1+l_2 \ri|^2} 
\ {\ov F}_{-2}(\ov \tau)\cr
&-\ {2\o  (T-\ov T)}\ T_2^2 
\int {d^2\tau \o \tau_2^4}\ 
\sum_{n_1,n_2 \atop  l_1,l_2} 
Q_R\ov Q_R\  e^{-2\pi i \ov T \det A}  e^{{-{\pi T_2} \o {\tau_2 U_2}}
\lf|n_1\tau+n_2U\tau -Ul_1+l_2 \ri|^2} 
\ {\ov F}_{-2}(\ov \tau)\ .}}
Again, for the integration we use the formulas of appendix B
to arrive at:
\eqn\fttufinal{\eqalign{
{\cal I}_{2,0}&=
8\pi^2 \p_T\lf(\p_T-{2\o T-\ov T}\ri)\lf(\p_U-{2\o U-\ov U}\ri)f
-{16\pi^2\o (T-\ov T)^2}\lf(\p_{\ov U}+{2\o U-\ov U}\ri)\ov f\cr
&=8\pi^2 f_{TTU}+2 G_T^{(1)}\ .}}
The second term in the integrand of \fttu\
might be identified with $4 G_T^{(1)}$, although a splitting of both terms
does not make sense because of modular invariance. Besides, only the
combination $8\pi^2 f_{TTU}+2 G_T^{(1)}$ can be written 
covariant w.r.t. \covsT.  

\subsubsec{4.2.3.\ $f_{TUU}$}

Finally, for the integral
\eqn\ftuu{
{\cal I}_{0,2}:={2\pi\o (U- \ov U)}\ \int d^2\tau
\  \sum_{(P_L,P_R)}\lf(P_L \ov P_R P_R^2-{P_L P_R\o \pi\tau_2}\ri) \   
{\hat Z}_{torus}(\tau,\ov\tau)\ {\ov F}_{-2}(\ov \tau)\ ,}
we borrow the results of section (4.2.2) and use mirror symmetry
$T\leftrightarrow U$ to obtain:
\eqn\ftuufinal{\eqalign{
{\cal I}_{0,2}&=
8\pi^2 \p_U\lf(\p_U-{2\o U-\ov U}\ri)\lf(\p_T-{2\o T-\ov T}\ri)f
-{16\pi^2\o (U-\ov U)^2}\lf(\p_{\ov T}+{2\o T-\ov T}\ri)\ov f\cr
&=8\pi^2 f_{TUU}+2 G_U^{(1)}\ .}}

\newsec{Six dimensional origin of the gauge couplings}

Let us consider the amplitudes discussed in the section 3 
from a more general point of view. The gauge kinetic
terms \egaugea\ are deduced from the Einstein term in six
dimensions upon dimensional reduction. In the Einstein
frame the latter does not receive any loop corrections neither in
$d=6$ nor in $d=4$. The relevant object to consider is a two graviton 
amplitude ($i=1,\ldots,6$)
\eqn\six{
\la :\eps_{ij}\bar\p X_1^i\Big[\p X_1^j+i(k_1\cdot\psi_1)\psi_1^j\Big]
e^{ik_1\cdot X_1}: 
:\eps_{kl}\bar\p X_2^k
\Big[\p X_2^l+i(k_2\cdot\psi_2)\psi_2^l\Big] e^{ik_2\cdot X_2}: \ra\ ,}
which may contain both ${\cal R}$ and 
${\cal R}_{ikjl}{\cal R}^{ikjl}$ corrections.
Here $\eps_{ij}$ is the gravitational polarization tensor in six 
dimensions.  
The amplitude is determined by expanding the elliptic genus 
${\cal A}$ of $K3$
w.r.t. ${\cal R}^2$. In general, in N=1,$d=6$ heterotic string theories 
only the $4$--form part of the elliptic genus gives rise to modular
invariant one--loop corrections \lsw\WL.
For the choice of instanton numbers
$(n_1,n_2)=(24,0)$ w.r.t. an $SU(2)$ gauge bundle which leads
to the gauge group $E_7\times E_8'$ this expansion is given by \lsw\WL\hm
\eqn\genus{
\lf. {\cal A}(\tau,{\cal F},{\cal R})\ri|_{4-form}\sim\lf[
({\cal R}^2-{\cal F}^2){\hat{\bar E}_2 \bar E_4\bar E_6 \o
\bar\eta^{24}}+{\bar E_6^2\o \bar\eta^{24}}
{\cal F}^2_{E_7}+{\bar E_4^3\o \bar\eta^{24}}{\cal F}^2_{E'_8}\ri]\ ,}
To saturate the fermionic zero modes one has to contract the four
fermions which is already of the order $\Oc(k^2)$.
Since the worldsheet integral of the bosonic contraction 
$\la \bar \p X^i_1\bar \p X_2^k\ra$ gives a zero result and thus the
$\Oc(k^2)$ term of the effective action vanishes, 
the next to leading order arises from contractions of $\bar \p X^i$ with
the exponential $e^{i k\cdot X}$ which contributes another $\Oc(k^2)$
term to the amplitude. For the $\Oc(k^4)$ contribution we thus obtain \WL:
\eqn\sixDelta{
\triangle_{{\cal R}^2}^{6d,N=1}\sim \int {d^2\tau \o \tau_2^2}
\lf(\bar E_2-{3\o \pi \tau_2}\ri) {\bar E_4 \bar E_6 \o \bar\eta^{24}}
=-8\pi\ .}
The $\hat{\bar E}_2$--piece arises from  
the worldsheet integral over the contractions of $6d$ space--time
bosons:
\eqn\contra{
\int d^2 z_{12}\la\ov\p\bar X_1^i X_2^m\ra\la\ov\p\bar X_2^k X_1^n\ra=
-\eta^{im}\eta^{kn}\int\Big[\bar\p
G_B(z_{12})\Big]^2=\eta^{im}\eta^{kn}{\pi^2\tau_2\o 8}\Big(\bar
E_2-{3\o\pi\tau_2}\Big)\ .}
The appearance of $\hat{\bar E}_2$ in \sixDelta\ may be also understood as
the gravitational charge
$Q_{\rm grav}^2=-2\p_\tau\ln\eta(\tau)=1/6\pi i\  E_2(\tau)$ \agnt.
After the torus compactification we obtain in $d=4$ \WL\agnt\hm
\eqn\fourDelta{
\triangle_{{\cal R}^2}^{4d,N=2}\sim\int {d^2\tau \o \tau_2}{Z}_{torus} 
\lf(\bar E_2-{3\o \pi \tau_2}\ri){\bar E_4 \bar E_6 \o \bar\eta^{24}}\ .}
There is however a subtlety w.r.t. to the indices $i,j,k,l$ 
in deducing the field theoretic kinematic content of \six\ 
for the four dimensional action:
Taking all $i,j,k,l$ as $d=4$ space--time indices $\mu,\nu$ gives
\sixDelta\ for the $\Rc_{\mu\nu\rho\sigma}\Rc^{\mu\nu\rho\sigma}$
correction in $4d$, i.e. \fourDelta\
after taking into account the zero modes of the internal torus.
However when we want to deduce the gauge kinetic term \stgauge,
we keep both $i$ and $k$ as internal indices $+$ which has been
defined in section 3 and 
extract the ${\cal O}(k^2)$--piece of \six\ [cf. \ig]:
\eqn\extract{
\int d^2 z_{12}\la \bar\p X_1^+\bar\p X_2^+\ra=
 2\pi^2 \tau_2 \bar{P}_R^2\ .}
This way we end up at
\eqn\fourF{\Delta_{(TT)}\sim\int {d^2\tau \o \tau_2}
\sum_{(P_L,P_R)}\hat{Z}_{torus} \bar P_R ^2 
{\bar E_4 \bar E_6 \o \bar\eta^{24}}\ .}
after taking into account all kinematic possibilities, in agreement
with \att.
Of course, the $\ov E_2$--part of \sixDelta, measuring the gravitational 
charge, does not occur in $\triangle_{(TT)}$.

\newsec{Conclusion}

We have calculated the one--loop threshold corrections
to the gauge couplings of $U(1)$ gauge bosons which
arise from heterotic N=2,\ $d=4$ compactifications on a torus.
These results fit into the framework
of the underlying N=2 supergravity theory.
Using (4.21) and \nsii\ we may cast the effective gauge couplings
\atu\ into the final form:

\eqn\finalgauge{
-{\triangle_{(TU)}\o 32\pi^2}={1\o 2} \re\lf\{ D_T D_U f\ri\}
={1\o 16\pi^2} \lf[\ln|j(T)-j(U)|^2-G^{(1)}+\sigma(T,U)\ri]\ ,}
where $\sigma(T,U)$ are the universal one--loop corrections
appearing in all gauge threshold corrections of heterotic
N=2 theories \nsii. The correction $G^{(1)}$ describes the
mixing of the dilaton and the moduli fields at one--loop \jv\dewit\nsii.

In section 4 we have calculated several world--sheet $\tau$--integrals
as they appear in string amplitude calculations from various
contractions of the internal bosonic coordinate fields.
These expressions appear quite general in 
heterotic torus compactifications from N=1 in $d=10,6,4$ to $d=8,4,2$.
The relevant string amplitudes take a generic form, given by a 
$\tau$--integral over the (new)
supersymmetric index \susyindex\ (or variants of it depending on $d$), 
completed with momentum insertions of internal
bosonic fields, which take into account either vertex operator contractions
or charge insertions. In the case of $K3\times T^2$ compactifications,
a part of the supersymmetric index \susyindex\ is related to a modular
function $f$, which is the N=2, $d=4$ prepotential. Many
$\tau$--integrals can be expressed through $f$ and its derivatives.
Such relations between string--amplitudes and a function $f$ and its
derivatives, as established here for N=2 in $d=4$, should also exist 
in any torus compactifications of e.g. $d=10,4$ heterotic string theory.

\smallskip

\ \ 
{\bf Acknowledgments:}
We are very grateful to I. Antoniadis, J.--P. Derendinger and W. Lerche 
for important discussions.  
K.F. thanks the Institut de Physique Th{\'e}orique de 
l'Universit{\'e} de Neuch{\^a}tel for the friendly hospitality.
This work is supported by the
Swiss National Science Foundation, and the European Commission TMR programme 
ERBFMRX--CT96--0045, in which K.F. is associated to HU Berlin and St. St.
to OFES no. 95.0856.

\appendix{A}{\bf Poisson resummation}

\def\bp {{\vec p }}  \def\by {{\vec y}} \def\bb {{\vec b}} \def\bc {{\vec c}} 

In this section we want to perform a Poisson transformation on
the expression
\eqn\start{
{\cal S}=\sum_{\bp\in \Lambda^\ast} e^{-\pi \bp^t \alpha \bp} 
e^{2\pi i \by^t \bp}(\bp^t A \bp+\bb^t \bp+a_0)(\bp^t D \bp+\bc^t \bp+e_0)\ ,}
for some matrices $\alpha,A,D$ ($\det\alpha\neq 0$), 
vectors $\by,\bb,\bc$ 
and scalars $a_0,e_0$.
This is achieved like one does a Fourier transformation on a periodic
function $F(\vec x)$, i.e. $[F(\vec x+\vec p_0)=F(\vec x)]$:
\eqn\fourier{
F(\vec x)=\sum_{\bp \in \Lambda} e^{-\pi (\bp+\vec x)^t\alpha(\bp+\vec
x)} e^{2\pi i \by^t(\bp+\vec x)}\ .}
With 
\eqn\dual{
F^{\ast}(\vec q)={1\o vol(\Lambda)}\int\limits_{-\infty}^\infty 
d \vec x\  e^{-2\pi i \vec q^t \vec x} F(\vec x)\ ,}
we may write:
\eqn\ffinal{\eqalign{
F(\vec x)&=\sum_{\vec q\in \Lambda^\ast}\  e^{2\pi i {\vec q}^t \vec x}
F^\ast(\vec q)\cr
&={1\o \sqrt{\det \alpha}}{1\o vol(\Lambda)}
\sum_{\vec q \in \Lambda^\ast} e^{-2\pi i {\vec q}^t \vec x}
e^{-\pi (\vec y+\vec q)^t\alpha^{-1}(\vec y+\vec q)}\ .}}
Along that way a Fourier transformation on \start\ yields

\eqn\resfour{\eqalign{
{\cal S}&={1\o \sqrt{\det \alpha}}{1\o vol(\Lambda^\ast)}
          \sum_{\vec q \in \Lambda} 
          e^{-\pi (\vec y+\vec q)^t\alpha^{-1}(\vec y+\vec q)}\ 
          \lf\{a'_0e'_0+{1\o 2\pi}\Big[e'_0 \tr(\alpha^{-1}A)
          +a'_0 \tr(\alpha^{-1}D)\ri.\cr
        &+\lf(\vec b+i(A^t+A)\alpha^{-1}(\vec y+\vec q)\ri)^t  \alpha^{-1}
          \Big(\vec c+i(D^t+D)\alpha^{-1}(\vec y+\vec q)\Big)  \Big]\cr
        &\lf. +{1\o 4\pi^2}\Big[\tr(\alpha^{-1}A)\tr(\alpha^{-1}D)+
          \tr(\alpha^{-1}A\alpha^{-1}D^t)+
           \tr(\alpha^{-1}A\alpha^{-1}D) \Big] \ri\}  }}
with the following abbreviations:
\eqn\abbrev{\eqalign{
a_0'&=a_0-(\vec y+\vec q)^t\alpha^\ast A\alpha^{-1}(\vec y+\vec q)+
i\vec b^t\alpha^{-1}(\vec y+\vec q)\cr
e_0'&=e_0-(\vec y+\vec q)^t\alpha^\ast D\alpha^{-1}(\vec y+\vec q)+
i\vec c^t\alpha^{-1}(\vec y+\vec q)\ .}}

\appendix{B}{\bf Integrals}

\subsec{Orbit $I_1$}

For the orbit $I_1$ we have to face the following integrals
\eqn\genericint{
I_1^{\alpha,\beta,n}:=\sum_{k,j,p}{\tilde I}_1^{\alpha,\beta,n}=
T_2\ e^{-2\pi i \ov Tkp}
\int\limits_{H_+} {d\tau \o \tau_2^{2+\beta}} \sum_{k,j,p}
\tau_1^\alpha e^{-{\pi T_2\o \tau_2 U_2}|k\tau-j-pU|^2}
e^{-2\pi i\ov\tau n}\ ,}
for the cases $\alpha=0,\ldots,4$ and $\beta=-1,0,\ldots,3$.
Clearly, the case $\alpha=0$ and $\beta=0$ corresponds to
the integral performed in \DKLII. The case $\beta=-1$ is needed for
the integrals appearing in sect. 4.2.
We expanded the anti--holomorphic function $\ov F(\ov \tau)$ in a power
series:
\eqn\series{
\ov F(\ov \tau)=\sum_{n \geq -1} c_n e^{-2\pi i\ov\tau n}\ .}
The $\tau_1$--integration of \genericint\ can be reduced to Gaussian
integrals:
\eqn\zwischen{\eqalign{
{\tilde I}_1^{\alpha,\beta,n}&={k^{-\alpha}\o k}\sqrt{T_2U_2}\ e^{-2\pi i \ov
Tkp}\ 
e^{2\pi T_2 kp}\ e^{-2\pi i {n \o k}(j+pU_1)}\cr
&\times\int_0^\infty {d\tau_2 \o 
\tau_2^{{3\o 2}+\beta}}\ e^{-{\pi T_2 \o U_2}(k+{nU_2 \o k T_2})^2\tau_2}
\ e^{-\pi p^2T_2 U_2/\tau_2}\ {\cal X}_\al\ ,}}
with: 
\eqn\intalpha{\eqalign{
{\cal X}_0&=1\cr
{\cal X}_1&=-i{n\o k} {\tau_2 U_2 \o T_2}+j+pU_1 \cr
{\cal X}_2&={1\o 2\pi}{\tau_2 U_2 \o T_2} + 
(-i{n\o k} {\tau_2U_2 \o T_2}+j+pU_1)^2\cr
{\cal X}_3&={3\o 2\pi} {\tau_2 U_2\o T_2}
(-i{n\o k} {\tau_2U_2 \o T_2}+j+pU_1)+
(-i{n\o k} {\tau_2U_2 \o T_2}+j+pU_1)^3\cr
{\cal X}_4&={3\o 4\pi^2} \lf({\tau_2 U_2\o T_2}\ri)^2+
{3\o\pi}{\tau_2 U_2\o T_2}(-i{n\o k} {\tau_2U_2 \o T_2}+j+pU_1)^2+
(-i{n\o k} {\tau_2U_2 \o T_2}+j+pU_1)^4\ .}}
Next, we have to do the $\tau_2$--integration. For $\beta=0,1,2$ 
we may borrow results\foot{We thank N.A. Obers for explanation of the notation
of \bachas. In their final formulae
one must replace $p\rightarrow |p|$.}
from \bachas\ for the integrals
\eqn\zwischenii{
{\tilde I}_1^{0,\beta,n}=\lf(k\o |p| U_2\ri)^\beta {\tilde I}_1 
                     \times \cases{\lf[1+n{{\cal U}_2\o{\cal
                     T}_2}\ri]^{-1} &$\beta=-1$\ ,\cr
                     1 &$\beta=0$\ ,\cr
                     \lf[1+{1\o {\cal T}_2}(n{\cal U}_2+{1\o 2\pi})\ri]
                     &$\beta=1$\ ,\cr
                     \lf[1+{1\o {\cal T}_2}(2n{\cal U}_2+{3 \o 2\pi})
                     +{1\o {\cal T}_2^2}(n^2{\cal U}_2^2+{3n{\cal U}_2
                     \o 2\pi}+{3 \o 4 \pi^2})\ri]&$\beta=2$\ , \cr
                     \lf[1+{1\o {\cal T}_2}(3n{\cal U}_2+{3 \o \pi})
                     +{1\o {\cal T}_2^2}(3n^2{\cal U}_2^2+{6n{\cal U}_2
                     \o \pi}+{15 \o 4 \pi^2})\ri.&$\ $\cr
                     \lf.\ \ \ \ +{1\o {\cal T}_2^3}(n^3{\cal U}_2^3+
                     {3n^2\o\pi}{\cal U}_2^2+{15\o 4 \pi^2}n {\cal U}_2+
                     {15\o 8\pi^3})\ri]&$\beta=3$\ , \cr}}
with
\eqn\zwischeniii{
\tilde I_1={1\o k|p|}e^{-2\pi i\ov T kp} e^{-2\pi i {n \o k}(j+pU_1-i|p|U_2)}
\ e^{2\pi T_2 k(p-|p|)}\ ,}
and: 
\eqn\moduli{
{\cal T}_2=k|p|T_2\ \ \ ,\ \ \ {\cal U}_2={|p|U_2 \o k}\ .}
Before we continue, let us recover 
\eqn\easy{\eqalign{
I_1^{0,0,n}&=\sum_{k>0 \atop l\in \IZ}
\delta_{n,kl}\ \Li_1\lf[e^{2\pi i(kT+lU)}\ri]+hc.\ , \cr
I_1^{0,1,n}&=\sum_{k>0 \atop l\in \IZ}
\delta_{n,kl}\ {\cal P}\lf[e^{2\pi i(kT+lU)}\ri]+hc.\ ,}}
with:
\eqn\calp{
{\cal P}(kT+lU)=\im(kT+lU)\ \Li_2\lf[e^{2\pi i(kT+lU)}\ri]+{1\o 2\pi}
\Li_3\lf[e^{2\pi i(kT+lU)}\ri]\ .}

Finally for the cases of interest, 
we have reduced everything to integrals
${\tilde I}_1^{0,\beta,n}$ given in eq. \zwischenii .
\eqn\last{
\kern-1em 
\eqalign{
k \tilde I_1^{1,\beta,n}&=-i{n\o k} {U_2 \o T_2}{\tilde I}_1^{0,\beta-1,n}
+(j+pU_1){\tilde I}_1^{0,\beta,n}\cr
k^2 {\tilde I}_1^{2,\beta,n}&=-{n^2\o k^2} {U_2^2 \o T_2^2}
{\tilde I}_1^{0,\beta-2,n}+{1\o 2\pi}{U_2 \o T_2}{\tilde I}_1^{0,\beta-1,n}
-2i(j+pU_1){n\o k} {U_2 \o T_2}{\tilde I}_1^{0,\beta-1,n}\cr
&\ \ \ +(j+pU_1)^2{\tilde I}_1^{0,\beta,n}\cr
k^3 {\tilde I}_1^{3,\beta,n}&=i{n^3\o k^3}{U_2^3\o T_2^3}
{\tilde I}_1^{0,\beta-3,n}-{3\o 2\pi}i{n\o k}{U_2^2\o T_2^2}
{\tilde I}_1^{0,\beta-2,n}-3{n^2\o k^2}(j+pU_1){U_2^2\o T_2^2}
{\tilde I}_1^{0,\beta-2,n}\cr
&\ \ \ +{3\o 2\pi}(j+pU_1)
{U_2\o T_2}{\tilde I}_1^{0,\beta-1,n}
-3i(j+pU_1)^2{n\o k}{U_2\o T_2}{\tilde I}_1^{0,\beta-1,n}+(j+pU_1)^3
{\tilde I}_1^{0,\beta,n}\cr
k^4 {\tilde I}_1^{4,\beta,n}&=
{n^4\o k^4}{U_2^4\o T_2^4}{\tilde I}_1^{0,\beta-4,n}
-{3\o \pi}{n^2\o k^2}{U_2^3\o T_2^3}{\tilde I}_1^{0,\beta-3,n}
+4i{n^3\o k^3}{U_2^3\o T_2^3}(j+pU_1){\tilde I}_1^{0,\beta-3,n}\cr
&\ \ \ -{6\o \pi}i{n\o k}(j+pU_1){U_2^2\o T_2^2}{\tilde
I}_1^{0,\beta-2,n}+{3\o 4\pi^2}{U_2^2\o T_2^2}{\tilde I}_1^{0,\beta-2,n}
-6{n^2\o k^2}{U_2^2 \o T_2^2}(j+pU_1)^2{\tilde I}_1^{0,\beta-2,n}\cr
&\ \ \ +{3\o \pi}{U_2\o T_2}(j+pU_1)^2{\tilde I}_1^{0,\beta-1,n}
-4i{n\o k}{U_2\o T_2}(j+pU_1)^3{\tilde I}_1^{0,\beta-1,n}
+(j+pU_1)^4{\tilde I}_1^{0,\beta,n}\ .}}
Now, the important and nice fact is, that after expanding the
expression
\last, the $j$--sum, in the combination of (4.10), (4.16), (4.25),
(4.32) and (4.34) becomes trivial and gives the restriction 
$n=kl\ ,\ l\in \IZ$.  
Then, after a straightforward calculation the orbits 
$I^{{\cal I}_{w_T,w_U}}_1$ belonging to the integral ${\cal
I}_{w_T,w_U}$ can be determined:
\eqn\IIZ{\eqalign{
I^{{\cal I}_{2,-2}}_1&=-\sum_{k>0\atop l\in \IZ}\sum_{p>0}
\delta_{n,kl}\lf(2{k^2\o p}+{k\o \pi T_2  p^2 }+{1 \o 4\pi^2 T_2^2 p^3 }\ri)
x^p\cr
&\ \ \ -\sum_{k>0\atop l\in \IZ}
\sum_{p>0}\delta_{n,kl}\ \lf(2 {U_2^2 l^2 \o T_2^2 p}+{U_2 l\o \pi T_2^2 p^2}+
{1\o 4\pi^2 T_2^2 p^3}\ri)\ov x^p\cr
I^{{\cal I}_{0,0}}_1&=-\sum_{k>0\atop l\in \IZ}\sum_{p>0}
\ \delta_{n,kl}\lf({2kl\o p}+{l\o \pi T_2 p^2}+
{k\o \pi U_2 p^2}+{1\o 2\pi^2 T_2 U_2 p^3}\ri)x^p+hc.\cr
I^{{\cal I}_{4,-2}}_1&=-2i\sum_{k>0\atop l\in \IZ}\sum_{p>0}\delta_{n,kl}\ 
k^3 x^p\cr
I^{{\cal I}_{2,0}}_1&=-\sum_{k>0\atop l\in \IZ}\sum_{p>0}\ \delta_{n,kl}
\lf(4i\pi k^2l+{2ik^2\o U_2 p}+{ik\o \pi T_2 U_2p^2}+{2ikl\o T_2 p}
+{i\o \pi^2 T_2^2 U_2p^3}+{il\o\pi T_2^2p^2}\ri)x^p\cr
&\ \ \ -\sum_{k>0\atop l\in \IZ}\sum_{p>0}\ \delta_{n,kl}
\lf({i\o 4\pi^2T_2^2 U_2p^3}-{il\o 2\pi T_2^2 p^2}\ri)\ov x^p\ ,}}
with $x:=e^{2\pi i(kT+lU)},\ \ov x:=e^{-2\pi i(k\ov
T+l\ov U)}$.

\subsec{Orbit $I_2$}

For the orbit $I_2$ the following integrals appear
\eqn\Ithree{
I^{\alpha,\beta,\gamma,n}_2=T_2\int_{-\h}^{+\h}d\tau_1{\int_0^\infty}
{d\tau_2 \o \tau_2^{2+\gamma}}\sum_{(j,p)}'j^\alpha p^\beta\ 
e^{-{\pi T_2 \o \tau_2 U_2}|j+Up|^2} e^{-2\pi i\ov\tau n}\ .}
Here the prime at the sum means that we do not sum over 
$(j,p)=(0,0)$, which is taken into account in $I_0$. The case
$\gamma=0$ describes the respective integral of \DKLII. In that case one 
has to regularize the integral. We will only need cases with
$\gamma\neq 0$. Therefore we have not to introduce an IR--regulator.
See also \kk\ for discussions. 
This also means that our results will not produce any non--harmonic
$\ln(T_2 U_2)$--pieces. Such terms are usually signals of potential 
anomalies arising in the IR.
The $\tau_1$--integration is trivial and simply projects onto massless
states, i.e. $n=0$. For the $\tau_2$--integration we again 
may use results of \bachas:
\eqn\intbach{
\int_0^\infty{d\tau_2\o \tau_2^{2+\gamma}} \sum_{(j,p)}'\ 
e^{-{\pi T_2 \o \tau_2 U_2}|j+Up|^2}=\Gamma(\gamma+1)\lf({U_2\o \pi
T_2}\ri)^{\gamma+1}\ \sum_{(j,p)}^{'}{1\o |j+p U|^{2\gamma+2}}\ .}
Therefore, we only have to concentrate on the sum
\eqn\sommer{
\sum_{(j,p)}'{j^\alpha p^\beta\o |j+Up|^{2\gamma+2}}\ .}

Let us first perform the summation for the case $p\neq 0$ and write:
\eqn\Zwisch{\eqalign{
\sum_{p\neq 0}{j^\alpha p^\beta\o
[(j+U_1p)^2+p^2U_2^2]^{\gamma+1}}&=
{1\o\gamma !}\sum_{p>0}\lf({1\o i} {\p\o \p\theta}\ri)^\alpha p^{\beta-2\gamma}
\lf({(-1)\o 2U_2} {\p \o \p U_2}\ri)^\gamma\cr
&\times\sum_{j=-\infty}^\infty\lf.\lf[{e^{i\theta j}\o (j+U_1p)^2+p^2U_2^2}+
{(-1)^\beta\ e^{i\theta j}\o (j-U_1p)^2+p^2U_2^2}\ri]
\ri|_{\theta=0}\ .}}
After using a Sommerfeld--Watson transformation, introduced in \hm, 
($C>0,\ 0\leq\theta\leq 2\pi$)
\eqn\watson{
\sum_{j=-\infty}^\infty{e^{i\theta j}\o (j+B)^2+C^2}={\pi \o C}e^{-i\theta
(B-iC)}{1\o 1-e^{-2\pi i(B-iC)}}+{\pi \o C}e^{-i\theta 
(B+iC)}{e^{2\pi i(B+iC)}\o 1-e^{-2\pi i(B+iC)}}\ ,}
we end up with
\eqn\ZW{\eqalign{
&\ {\pi\o \gamma!}\lf({-1\o 2 U_2} {\p \o \p U_2}\ri)^\gamma\Big\{
{1\o U_2}\Big[[(-U)^\alpha+(-1)^\beta U^\alpha]\sum_{l>0}
\Li_{1-\alpha-\beta+2\gamma}(q_U^l)\cr
&\ \ \ +[(-\ov U)^\alpha+(-1)^\beta \ov U^\alpha]
\sum_{l>0}\Li_{1-\alpha-\beta+2\gamma}(\ov q_U^l)\cr
&\ \ \ +[(-\ov U)^\alpha+(-1)^\beta U^\alpha]
\zeta(1-\alpha-\beta+2\gamma)\Big]\Big\}\ .}}

Let us now come to the case $p=0$ of \sommer
\eqn\asdf{
Q^{\alpha,\beta,\gamma}:=\delta_{\beta,0}\ [1+(-1)^\alpha]
\ \sum_{j=1}^\infty {1\o j^{2\gamma+2-\alpha}}\ ,}
which only gives a non--zero contribution for $\beta=0$. 
Moreover, we must only consider $\alpha \in 2\IZ$.
For the examples we discuss in section 4.1. we have $\alpha+\beta=\gamma=2$ 
and for the cases in section 4.2, $\alpha+\beta=4,\ \gamma=3$.
In both cases, the sum \asdf\ becomes:
\eqn\cont{
Q^{2,0,2}=Q^{4,0,3}=2\sum_{j=1}^\infty {1\o j^4}={\pi^4\o 45}\ .}

Putting everything together, we obtain for $I_2^{\alpha,\beta,\gamma,n}$:
\eqn\FINAL{\eqalign{
I_2^{\alpha,\beta,\gamma,n}&=\delta_{n,0}T_2\ \Gamma(\gamma+1)
\lf({U_2\o \pi T_2}\ri)^{\gamma+1}\cr
&\times\Big\{\ {\pi\o \gamma!}\lf({-1\o 2 U_2} {\p \o \p U_2}\ri)^\gamma\Big[
{1\o U_2}\Big([(-U)^\alpha+(-1)^\beta U^\alpha]\sum_{l>0}
\Li_{1-\alpha-\beta+2\gamma}(q_U^l)\cr
&\ \ \ +[(-\ov U)^\alpha+(-1)^\beta \ov U^\alpha]
\sum_{l>0}\Li_{1-\alpha-\beta+2\gamma}(\ov q_U^l)\cr
&\ \ \ +[(-\ov U)^\alpha+(-1)^\beta U^\alpha]
\zeta(1-\alpha-\beta+2\gamma)\Big)\Big]\cr  
&+Q^{\alpha,\beta,\gamma}\Big\}\ .}}

\listrefs
\end